\documentclass[sigconf]{aamas} 
\usepackage{balance} 

\usepackage{mathtools}          
\mathtoolsset{showonlyrefs}     
\usepackage{graphicx}           
\usepackage{subcaption}         
\usepackage{makecell}           
\usepackage[space]{grffile}     
\usepackage{url}                
\usepackage{enumitem,kantlipsum}
\usepackage{tikz}
\usepackage{circuitikz}
\usetikzlibrary{intersections, positioning}
\usepackage{multirow}
\let\classAND\AND
\let\AND\relax
\usepackage{algorithm}
\usepackage{algpseudocode}

\let\AND\classAND
\AtBeginEnvironment{algorithmic}{\let\AND\algoAND}

\usepackage{xcolor} 

\makeatletter
\gdef\@copyrightpermission{
  \begin{minipage}{0.2\columnwidth}
   \href{https://creativecommons.org/licenses/by/4.0/}{\includegraphics[width=0.90\textwidth]{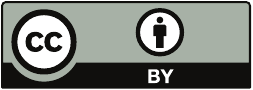}}
  \end{minipage}\hfill
  \begin{minipage}{0.8\columnwidth}
   \href{https://creativecommons.org/licenses/by/4.0/}{This work is licensed under a Creative Commons Attribution International 4.0 License.}
  \end{minipage}
  \vspace{5pt}
}
\makeatother

\setcopyright{ifaamas}
\acmConference[AAMAS '25]{Proc.\@ of the 24th International Conference
on Autonomous Agents and Multiagent Systems (AAMAS 2025)}{May 19 -- 23, 2025}
{Detroit, Michigan, USA}{Y.~Vorobeychik, S.~Das, A.~Nowé  (eds.)}
\copyrightyear{2025}
\acmYear{2025}
\acmDOI{}
\acmPrice{}
\acmISBN{}

\acmSubmissionID{609}

\title{Combining Planning and Reinforcement Learning for Solving Relational Multiagent Domains}

\author{Nikhilesh Prabhakar\footnotemark[1]}
\affiliation{
    \institution{The University of Texas at Dallas}
    \city{Richardson}
    \state{Texas}
    \country {USA}
}
\email{nikhilesh.prabhakar@utdallas.edu }

\author{Ranveer Singh\footnotemark[1]}
\affiliation{
  \institution{The University of Texas at Dallas}
  \city{Richardson}
  \state{Texas}
  \country {USA}
}
\email{ranveer.singh@utdallas.edu}
\author{
    Harsha Kokel
    }
\affiliation{
    \institution{IBM Research}
    \city{San Jose}
    \state{California}
    \country {USA}
}
\email{harsha.kokel@ibm.com}

\author{
    Sriraam Natarajan
    }
\affiliation{
    \institution{The University of Texas at Dallas }
    \city{Richardson}
    \state{Texas}
    \country{USA}\
    \email{sriraam.natarajan@utdallas.edu}
}
\email{sriraam.natarajan@utdallas.edu}

\author{
    Prasad Tadepalli
    }
\affiliation{
    \institution{Oregon State University}
    \city{Corvallis}
    \state{Oregon}
    \country {USA}
}
\email{tadepall@eecs.oregonstate.edu}

\begin{abstract}
Multiagent Reinforcement Learning (MARL) poses significant challenges due to the exponential growth of state and action spaces and the non-stationary nature of multiagent environments. This results in notable sample inefficiency and hinders generalization across diverse tasks. The complexity is further pronounced in relational settings, where domain knowledge is crucial but often underutilized by existing MARL algorithms. To overcome these hurdles, we propose integrating relational planners as centralized controllers with efficient state abstractions and reinforcement learning. This approach proves to be sample-efficient and facilitates effective task transfer and generalization.
\end{abstract}


\newcommand{\BibTeX}{\rm B\kern-.05em{\sc i\kern-.025em b}\kern-.08em\TeX}

\begin{document}


\pagestyle{fancy}
\fancyhead{}


\maketitle 

\footnotetext[1]{*Equal contribution}
\section{Introduction}
\label{sec:introduction}
Building multiple agents capable of learning to reason and act under uncertainty in large and complex environments has long been a cherished goal of AI. 
Reinforcement learning (RL) ~\citep{intro_rl_sutton_barto} and multiagent RL~\citep{MARLBook} techniques have long been developed for learning under uncertainty and in the presence of multiple agents, respectively. Several previous research efforts have extended these methods to hierarchical domains 
with multiple levels of state and action abstractions~\citep{maxq, hams, options}.

Statistical Relational Learning and AI (StaRAI)~\citep{SRLBook, StaRAIBook}, on the other hand, have dealt with learning in the presence of varying numbers of objects and relations, i.e., in relational domains. However, relational RL~\citep{rrl} is relatively unexplored, and while some methods exist~\citep{van2012rrlsurvey}, they do not scale for large tasks and are certainly not easily extensible to multiagent settings. A promising direction is exploiting the combination of hierarchical (and relational) planning to explore multiple levels of abstraction and RL to learn lower-level policies \citep{kokel2021reprel, TaskableRL}.

Inspired by the success in these different sub-areas of AI, we propose a method that leverages the power of a relational hierarchical planner to act as a centralized controller for multiagent learning in noisy, relational domains. Our proposed approach, called {\em Multiagent Relational Planning and Reinforcement Learning (MaRePReL)}, uses planning for task decomposition, centralized control, and agent allocation, StaRAI for constructing task-specific representations, and deep RL for effective and efficient learning with these specialized representations. 

We make the following key contributions: (1) As far as we are aware, we present the first multiagent system for relational multiagent domains that can generalize across multiple objects and relations. As we show in the related work, significant literature exists in multiagent systems, relational learning, and the integration of planning and learning. Ours is the first work to combine all these directions in the context of multiagent systems.
(2) To achieve this, we develop MaRePReL, an integrated planning and learning architecture capable of multiagent learning under uncertainty in relational domains. Specifically, MaRePReL's effective learning and reasoning power stems from its representation of relational information, the decomposition of higher-level plans, and the use of deep RL at the lowest level. (3) Finally, we demonstrate our approach's effectiveness and generalization abilities in a few relational multiagent domains. We compare against different deep RL based multiagent baselines, including one that explicitly uses the sub-task information, and illustrate the superiority of our approach.

The rest of the paper is organized as follows: after reviewing the related work and presenting the necessary background, we outline our multiagent framework and discuss the algorithm in greater detail. We then present the experimental evaluation on a few relational multiagent domains before concluding the paper by discussing the areas of future research.

\section{Related Work}
\begin{figure}
\centering
\resizebox{1\linewidth}{!}{
\begin{circuitikz}
\tikzstyle{every node}=[font=\normalsize]
\draw  (6.25,13.5) circle (0cm);
\draw [ color={rgb,255:red,153; green,193; blue,241} , fill={rgb,255:red,153; green,193; blue,241}, opacity=0.4] (7.5,15.75) circle (3.25cm) ;
\draw [ color={rgb,255:red,249; green,240; blue,107} , fill={rgb,255:red,249; green,240; blue,107}, opacity=0.4] (10,12.75) circle (3cm);
\draw [ color={rgb,255:red,143; green,240; blue,164} , fill={rgb,255:red,143; green,240; blue,164}, opacity=0.4] (5.25,12.75) circle (3cm);
\node [font=\normalsize] at (7.5,16.25) {\textbf{MARL}};
\node [font=\normalsize] at (5.25,13) {\textbf{HRL}};
\node [font=\normalsize] at (10,13) {\textbf{RRL}};
\node [font=\normalsize] at (3.75,14) {MaxQ \citep{maxq}};
\node [font=\normalsize] at (3.75,13) {Options \citep{options}};
\node [font=\normalsize] at (5.25,11.5) {TaskableRL \citep{TaskableRL}};
\node [font=\normalsize] at (11.75,14.5) {RRL \citep{rrl}};
\node [font=\normalsize] at (10,15.5) {MARRL \citep{marrl}};
\node [font=\normalsize] at (11.25,12.25) {Rex-D \citep{rrl_guidance}};
\node [font=\normalsize] at (11.25,11.25) {Fitted Q \citep{DasFittedQ}};
\node [font=\normalsize] at (5.1,15.25) {HMARL \citep{hmarl}};
\node [font=\normalsize] at (9.5,14.75) {MARL-DILP \citep{marl_ilp}};
\node [font=\normalsize] at (6,14.75) {HSD \citep{hsd}};
\node [font=\normalsize] at (6,13.5) {ALMA \citep{alma}};
\node [font=\normalsize] at (5.5,14) {Haven \citep{haven}};
\node [font=\normalsize] at (7.75,12.3) {RePReL \citep{kokel2021reprel}};
\node [font=\normalsize] at (7.75,11.9) {HRRL \citep{tadepalli_hrrl}};
\node [font=\normalsize] at (7.5,13.75) {MaRePReL};
\node [font=\normalsize] at (8.25,17.5) {MADDPG \citep{maddpg}};
\node [font=\normalsize] at (5.5,17) {Q-FTRL \citep{sampleefficientMARL}};
\node [font=\normalsize] at (9.25,17) {M-QMIX \citep{sampleGene}};
\node [font=\normalsize] at (6,17.75) {QMIX \citep{Qmix}};
\end{circuitikz}
}%
\caption{Our proposed framework w.r.t existing literature on relational, hierarchical, and multiagent RL}
\Description[]{Related Literature for our work}
\label{fig:Venn}
\end{figure}
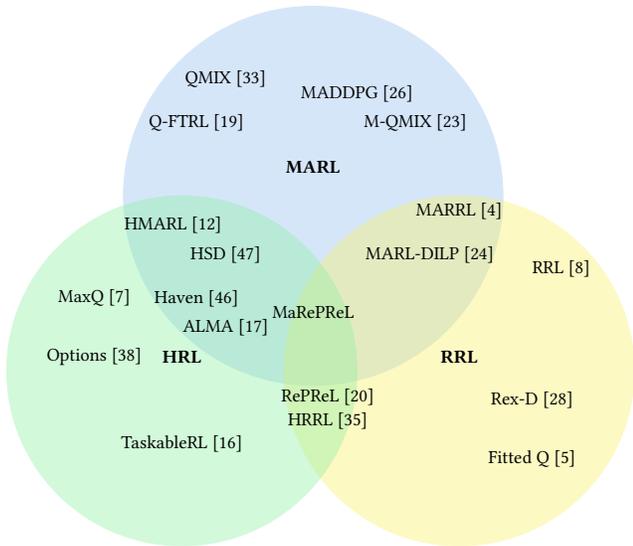

Research in RL in the past three decades has focused on several extensions that make them adaptable to several real-world scenarios. First, Hierarchical Reinforcement Learning (HRL) methods have been introduced to manage complex tasks by decomposing them into smaller, more manageable subtasks \citep{hams}. These allow for more efficient learning and problem-solving by utilizing multiple levels of abstraction. Second, Relational Reinforcement Learning (RRL) addresses the complexity of environments where states and actions consist of objects and the relationships between those objects \citep{rrl}. RRL exploits a higher-order representation of the underlying relational structure to improve learning in such domains. Third, Multi-Agent Reinforcement Learning (MARL) has been developed to handle environments where dynamic changes arise from the presence and actions of other agents, making it particularly useful in competitive or cooperative multi-agent settings \citep{MARLBook}. Before introducing our framework, which addresses all three challenges, namely hierarchies, relational structures, and multi-agent domains, we review the relevant literature for these three RL extensions.

\textbf{Hierarchical Reinforcement Learning (HRL)} algorithms have been developed to tackle the complexity of long-horizon tasks by breaking them down into smaller, more manageable subtasks. Frameworks such as the Options  \citep{options} and MAXQ \citep{maxq} facilitate the learning of hierarchical policies across multiple levels of abstraction. By exploiting temporal abstraction, HRL transforms the original long-horizon task into a sequence of shorter-horizon subtasks, where each subtask represents a high-level action that spans a longer period than the lower-level actions carried out by agents deeper in the hierarchy. This hierarchical structure enhances the agent’s ability to operate effectively over extended time horizons and significantly improves learning efficiency \citep{hrl_survey}. 

\textbf{Multiagent Reinforcement Learning (MARL)} extends reinforcement learning to systems with multiple agents, where they interact with the environment to maximize cumulative rewards \citep{MARLBook}. However, MARL introduces its own unique set of challenges. The first challenge is the curse of dimensionality, where the increasing number of agents leads to an exponential increase in the sizes of state and action spaces. The second challenge is the non-stationary nature of the environment 
due to the actions taken by the other agents. The final issue for MARL is the sample inefficiency due to the large amount of data required to train such agents \citep{zhang2021multi}.

Numerous solutions have been proposed to address the challenges outlined above, falling into two main categories: those adapting the underlying architectures for the RL agents and those considering the overall tasks performed by the agents. In the former category, methods use function approximation techniques to combat the curse of dimensionality \citep{dreduction}. In addition, Centralized Training, and Decentralized Execution (CTDE) methods such as QMIX and MADDPG address the non-stationary nature of the environment \citep{Qmix, maddpg}. Generative modeling or mask reconstruction algorithms \citep{sampleGene, sampleefficientMARL} also fall in this category. In the latter category, hierarchical approaches such as HMARL \citep{hmarl} and HSD \citep{hsd} utilize task decomposition and hierarchical structures in multiagent settings to define task abstractions and improve sample inefficiency by filtering out irrelevant parts of the state space. Additionally, the structured task hierarchies introduced in such methods can facilitate agent communication to address the non-stationary nature of different multiagent environments. While powerful in standard propositional (and continuous) settings, these methods do not address the challenge of a rich, relational structure in the environment.

\textbf{Relational Reinforcement Learning (RRL)} considers the task of learning in environments where states and actions involve relationships between objects and their properties, i.e., relational domains \citep{rrl}. In these domains, RL agents must explicitly learn to reason about and exploit the relationships between objects ~\citep{tadepalli2004relational}. 
Previously, several works have demonstrated the need for a rich relational representation to be explicitly used inside the learning algorithms as against simply grounding all the objects and obtaining a feature-based representation~\cite{StaRAIBook,liftedbook}. A key advantage of relational representations is their ability to support abstractions and facilitate generalization and effective transfer across tasks \citep{SRLBook, starai_nesy, rrl_survey_ottolo}.  However, finding optimal policies in many relational domains is intractable even for moderately large problems \citep{tadepalli2004relational}. To mitigate this issue, algorithms that incorporate guidance and domain knowledge as constraints on the policy space have been developed ~\citep{rrl_guidance}. 

\textbf{Planning and RL integration} have been explored to exploit the power of hierarchical planning with deep RL enabling the use of HRL in continuous domains. While Taskable RL \citep{TaskableRL} demonstrated significant performance improvement, the underlying planner was still propositional, limiting their applicability to relational problems with varying numbers of objects and relations.

An ideal RL learning algorithm should be able to not only handle the rich relational structure of the domain but also have the ability to represent and reason with the decomposition of complex tasks into smaller ones. In other words, the algorithm must be capable of representing and reasoning with both hierarchies and relational structures. 
One such recent framework, RePReL \citep{kokel2021reprel}, employs a hierarchical relational planner to implement task-specific policies and uses Deep RL to work on hybrid relational domains~\citep{kokel2021reprel,kokel2022Journal}. To interface the higher-level planner with the Deep RL, a {\em hand-crafted} abstract reasoner is employed to lift the reasoning process and construct smaller lower-level Markov Decision Processes (MDPs) that can be solved efficiently. This approach has been demonstrated to be successful in domains with varying numbers of objects, complex task structures, and continuous state-action spaces. 

While the RePReL framework successfully handled relations and hierarchies in continuous spaces, it can not handle multiagent systems. More precisely, given the three-pronged challenge of complex task structures, rich object-centric environments, and multiagent 
domains, several advances have been made in each of these specific directions. Also, in the recent past, methods that arise from the combinations of these methods -- for instance, HRL with RRL~\citep{tadepalli_hrrl, maxq, kokel2021reprel}, MARL with RRL~\citep{marrl, marl_ilp}, HRL with MARL~\citep{hmarl, hsd, haven, alma} -- have been proposed. However, no significant research encompasses all three of these challenges (see Figure~\ref{fig:Venn}).

It is precisely this gap that we aim to address in this work. Specifically, we extend the RePReL framework to multiagent settings, utilizing the planner as both a scheduler and a centralized controller. Unlike RePReL, where the planner is solely responsible for task decomposition, our proposed framework also distributes tasks among multiple agents. This key enhancement enables our approach to effectively solve relational multiagent domains, as we explain in the next section.

\section{Multiagent Relational Planning and Reinforcement Learning (MaRePReL)}
\label{sec:mareprel}
We consider the problem of coordinating multiple agents to solve continuous, relational problems. We will first provide a high-level overview of our framework, MaRePReL, combining relational and hierarchical planning with deep reinforcement learning before defining the problem formally. MaRePReL employs the following: 
\begin{enumerate}
    \item \textbf{Planner as controller:}
   One of our key contributions is to view the (relational) planner as a centralized controller that obtains the current state as input and creates a set of agent-specific plans. In the spirit of two-level systems, the planner not only decomposes the tasks into subtasks but also assigns the subtasks to appropriate agents. 
   To facilitate this, the controller consists of two specific components:
    \begin{enumerate}
        \item \textbf{Relational Hierarchical Planner:} Recall that the goal is to decompose tasks in the presence of varying numbers of objects and relations between them. Consequently, the first sub-component is a relational,
        hierarchical planner, which uses a first-order representation to model the objects and relationships in the domain. As one can view hierarchies as a specific form of relations, this planner can decompose the goals into a temporally ordered series of subgoals. 
        \item \textbf{Task Distributor: } The planner output is typically the task decomposition and does not bind the tasks to the specific agents. We use a task distributor as part of the relational planner to divide the ordered plan provided into agent-specific sub-plans using agent constraints for the different tasks.
    \end{enumerate}
\item \textbf{Abstraction Reasoner: } Following the previous work on single-agent learning (RePReL), we use Dynamic-First Order Conditional Influence (D-FOCI) statements~\citep{dfoci}
to capture domain knowledge that is then used to reason and construct the relevant parts of the state space that the lower-level RL agents use. This step at the outset is similar to RePReL, where the inference step is still hand-crafted, and leveraging lifted inference~\citep{liftedbook} to perform this step automatically remains a future direction. However, it must be emphasized that this step is more difficult in the multiagent setting as the optimal allocation of tasks to agents requires considering all agent states. Hence, typical single-agent-based abstraction reasoners do not suffice for this setting as they do not capture the true optimal value functions.

\item \textbf{Multiple Deep RL agents:} Given the current subtask from the planner, the corresponding (deep) RL agent identified by the task distributor learns a generalizable, task-specific policy. Assuming that the abstraction reasoner identifies the relevant part of the state space, learning is both effective and efficient with the additional advantage of being generalizable since the learned policies can be shared among multiple agents (as shown in our experiments). 

\end{enumerate} 

\begin{figure*}
    \Description[The architecture layout]{}
    \begin{subfigure}{0.48\linewidth}
        \includegraphics[scale=0.7]{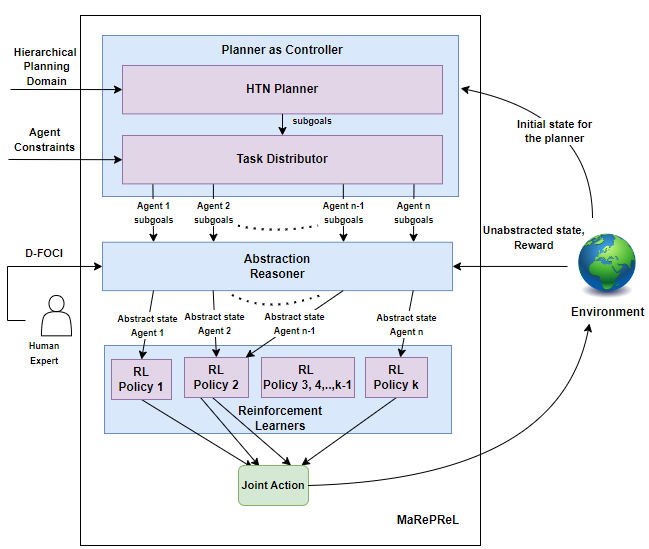}
    \end{subfigure}
    \begin{subfigure}{0.48\linewidth}
     \includegraphics[scale=0.12]{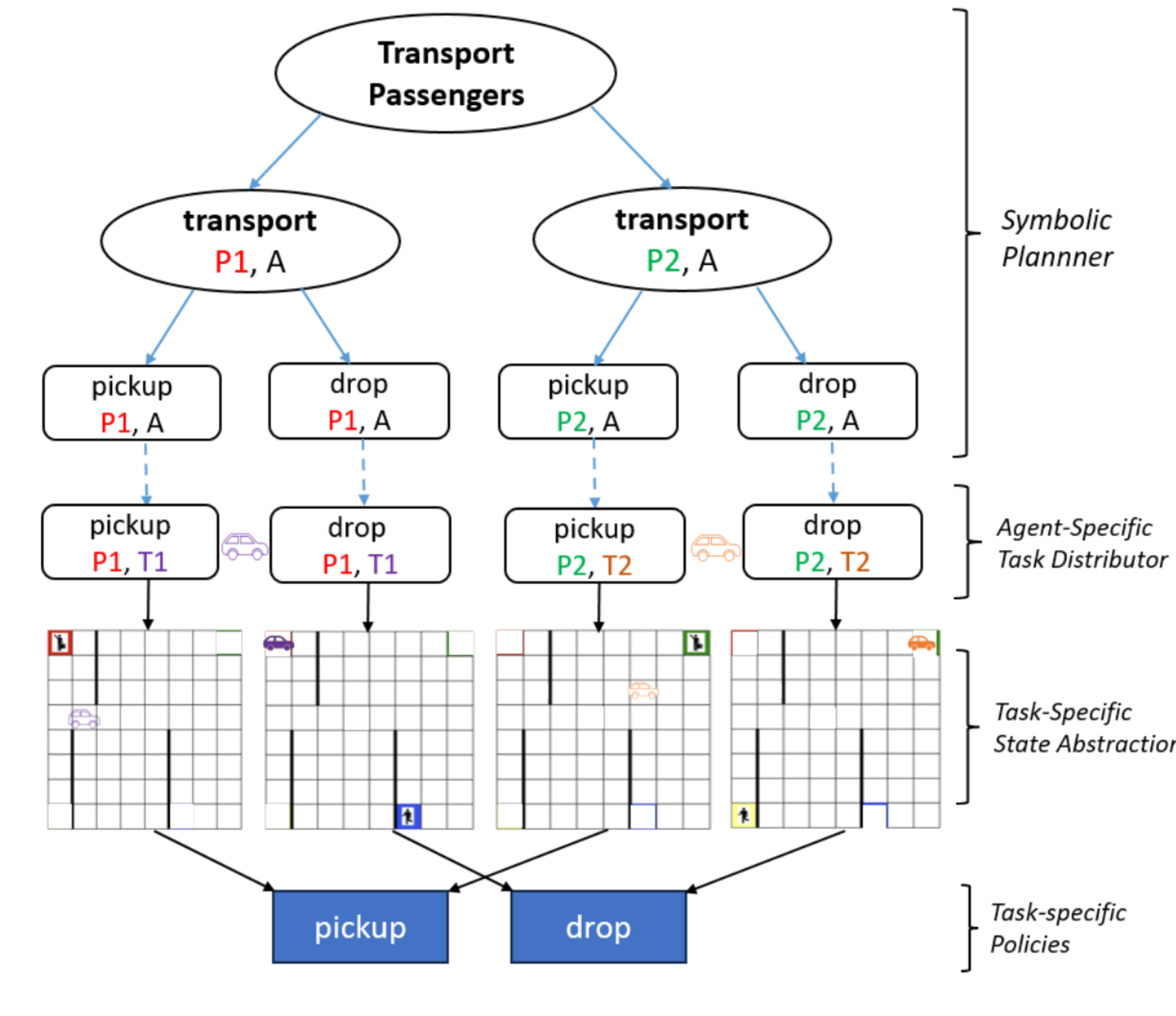}  
    \end{subfigure}
    \caption{MaRePReL architecture and application in the taxi world for the task of transporting two passengers}
    \label{fig:mareprel_arch}
\end{figure*}

The broad overview of our proposed approach is presented in Figure \ref{fig:mareprel_arch}. The planner decomposes the higher-level tasks into appropriate lower-level tasks using a relational representation of the current state and lifted operators. The distributor identifies the appropriate RL agent for the current subtask, thus making this combination an effective centralized controller. Given the subtask and the current (abstract/relational) state, the abstraction reasoner selects a smaller state representation by identifying the appropriate parts of the state space that are relevant to the current subtask. Finally, the RL agent either learns the policy or executes the ones it already learned (for instance, agent $A1$ might have learned the {\em pickup} subtask that can be used directly by agent $A2$ to execute this specific subtask). Note that while our experiments assume all the agents to be homogenous, this is not a necessary assumption for our formulation, where specific constraints can be used to allocate the tasks accordingly. 

\subsection{Problem Formulation}
While one could envision using relational partially observable MDPs (RPOMDPs) to model the problem with the current task being the hidden component, issues arise when modeling it as one. First, the abstraction reasoner has to track the current hidden state actively. Second, the use of decentralized RPOMDPs \citep{wang2010rpomdp} requires creating multiple RMDPs one for each task-agent formalism which would, in general, be larger than the smaller RMDPs created by our formalism using abstractions. Finally, for larger agent-task combinations, the reasoning over belief states requires approximate probabilistic inference over relational states and lifted probabilistic inference, which is outside the current work's scope. Therefore, we consider modeling the problem using Markov games ~\citep{littman1994markov}. We build upon the framework of relational Markov games ~\citep{rmg}, extending it to handle goal-oriented problems. We formalize this extension as a goal-directed relational Markov game (GRMG), defined as follows:

\textbf{Definition 1:} \emph{A goal-directed relational Markov Game (GRMG)
is represented as $M= \langle N, S, A^i_{i \in N}, P, R^i_{i \in N}, \gamma, G \rangle$ where $N$ is the number of agents, $S$ is the set of (relational) states, $A^i$ is the set of actions for the $i^{th}$ agent, and $A \coloneqq A^1 \times A^2 \times .....  \times A^N$ is the joint action space, $P = Pr(s'|s, a)$ is the transition probability function for transitioning from $s$ to $s'$ where $s, s' \in S$ and $a \in A$, $R^i = S \times A \times S \rightarrow  \mathbb{R}$ (set of real numbers) is the reward function for the $i^{th}$ agent, representing the instantaneous reward received by the agent on transitioning from one state to next after taking action, $\gamma \in [0, 1)$ is the discount factor, $G$ is the set of goals, the agents need to achieve. The states $S$ and actions $A$ are defined by the set of objects $E$, predicates $Q$, and action types $Y$.}

MaRePReL solves GRMGs using a combination of multiagent planning and RL, as shown in Figure~\ref{fig:mareprel_arch}.  A problem instance for a GRMG is defined similarly to in GRMDP~\citep{kokel2021reprel} as a pair $(s \in S, g \in G)$, where $s$ and $g$ is the initial state and the goal (a partial state), defined by a set of positive or negative literals. 
The success probability of a joint policy (say $\pi$) is the probability of all the goals $g \in G$ being achieved under the policy $\pi$. The expected utility of the joint policy is the expected total discounted reward before the policy terminates, either due to the successful completion of goals or timing out from exceeding the maximum possible length of the trajectory.

\subsection{Relational Planner}
The environment's state can be represented as an abstract planning problem using a planning description language \citep{holler2020hddl}. The hybrid planning domain $D = \langle Q, O, C, M \rangle$,  consists of a set of predicates $Q$ that describes the current state, a finite set of operators $O$ which are the high-level actions executable by the agents, a set of ordering constraints $C$ that is necessary to construct a consistent plan, and methods $M$ that can decompose the goal set into an ordered sequence of operators. A multiagent planning (MAP) problem can be defined as follows:

\textbf{Definition 2} \emph{For a given domain D, a Multiagent Planning (MAP) problem $P = \langle D, S, G, AG \rangle $, consists of the initial state of the problem $S$, the set of goals $G$ that need to be completed, and a group of agents $AG$ that need to coordinate together to reach the goal state. }

For the above MAP problem, the planner plays a crucial role by controlling the tasks performed by each agent. It maps the target set of goals $G$ into a set of grounded task-specific operators $O$ and distributes them to different agents. Hierarchical Task Network (HTN) planners such as SHOP \citep{shop} can be used to generate a total order plan for a given instance of the environment. The grounded plan (high-level plan) and a set of ordering constraints are used to distribute the tasks to create agent-specific plans (sub-plans). A greedy approach is used to schedule tasks that involve forward chaining \citep{kvarnstrom2011planning}. It examines the causal links between operators and prevents tasks from being assigned to agents that cannot execute them. The causal links $L = (l_1, l_2, \cdots, l_n)$ define a partial ordering between operations. Each link is of the form $ (O_p, \textit{eff}, O_q) $ where $\textit{eff}$ is the effect of completing task $O_p$ and one of the preconditions for task $O_q$ \citep{weld1994pocl}. Our task distributor ensures that the same agent performs the causally linked operations. 

For each agent $a$, the task distributor returns a partial plan  $\Pi^a = [o_1, o_2, o_3, ..., o_n]$ where $o$ is an operator with $I(o)$ being the precondition of the operator and $\beta(o)$ being the necessary effects of the operator. Since the operators only consider the action space of the agent currently using them, and the operators are shared among the different agents, all working on the same underlying environment,  we can define the sub-goal RMDP $\mathcal{M}_o$ for each operator to solve the problem like in RePReL \citep{kokel2021reprel}. 
\begin{algorithm}[!t]
    \caption{MaRePReL algorithm}
    \label{alg:MaRePReL}
    \begin{algorithmic}[1]
    \Statex \textbf{Input:} Multiagent Planner $\mathcal{P}$, Operators $O$, Agents $A$,  goal set  $g$,  D-FOCI statements $F$, num of iterations $i$, num of episodes per iteration $k$, batch size $b$, terminal reward $t_R$
    \Statex     \textbf{Output:} RL policies $\pi = \{\pi_o| \forall o \in O\}$
    \State Initialize the RL policies $\pi = \{\pi_o|\forall o \in O\}$ and buffers $\mathcal{D} = \{\mathcal{D}_o | \forall o \in O\}$
    \For{iteration $\in i$}
    \For{episode $\in k$}
        \State $s \gets $ starting state of the environment 
        \State $\Pi \gets \mathcal{P}(s, g)$ 
        \State $\phi \gets$ Pop the first task for each agent from $\Pi$
        \While{$\phi$ is not empty}
        \State $actions \gets$ \textbf{GetAgentActions($s, \phi, \pi, F, A)$}
        \State $s, \mathcal{D}, \phi, \textit{PlanValid}$  $\gets$
        
        \quad \quad \quad \textbf{RePReLStep}$(s, \textit{actions}, \mathcal{D}, t_R, \phi, \Pi, F, A)$
        \If{not $PlanValid$ }
           \State  $\Pi \gets \mathcal{P}(s, g)$ \Comment{Recompute the plan}
           \State $\phi \gets$ Pop the first task for each agent in $\Pi$
        \EndIf
        \EndWhile
    \EndFor
    \For{each operator $o \in O$}
    \State Sample batch $\mathcal{D}_b$ from the corresponding buffer $\mathcal{D}_o$
    \State Update Policy $\pi_o$ using the buffer $\mathcal{D}_b$
    \EndFor
    \EndFor
    \State \textbf{return} $\pi$
    \end{algorithmic}
\end{algorithm}
\subsection{Task-specific Abstraction}
While the planner decomposes the task and the task distributor identifies the appropriate agent, the resulting state space can still be prohibitively expensive for effective learning. Consequently, the abstraction reasoner becomes crucial in constructing a smaller state space.  In GRMG, states are represented as conjunctions of literals. Like RePReL, prior knowledge that describes the relation between rewards, and sub-goals is then described using an extended First Order Conditional Influence statements (FOCI) \citep{dfoci} called Dynamic FOCI (D-FOCI) statements. D-FOCI statements, represented by an example below, are the first-order language rules used to specify the direct conditional influences between literals in the domain. The rules defined over the predicates by a domain expert express the relation between domain predicates at a different time step.
 \begin{equation*} 
    pickup(P, T): \{ taxi(T, L1), at(P, L) \} \xrightarrow{+1} in\_taxi(P, T) 
 \end{equation*} 
The above rule states that when executing a task $pickup(P, T)$, only the location $L1$ of taxi $T$ and the pickup location $L$ of passenger $P$ influence the state predicate $in\_taxi$. The relational planner provides agent-specific plans that contain the grounded operators. Substituting the variables grounded by our planner in the D-FOCI statements will provide us with the set of literals on which our task-based RL policies can be trained. If the sub-plan for agent $t1$ contains the grounded operator $pickup(p1, t1)$, we can use the substitution $\theta = \{P/p1, T/t1, L/r, L1/l1\}$ to get the grounding,
\begin{equation*}
pickup(p1, t1) : \{ taxi(t1, l1), at(p1, r) \}  \xrightarrow{+1} in\_taxi(p1,t1)
\end{equation*}
which provides us with information on the relevant state literals for the task $pickup(p1,t1)$ as $taxi(t1,l1), at(p, r)$, and $in\_taxi(p1,t1)$. This implies that the task of picking up $p1$, when assigned to $t1$, only needs the information above, and the locations and in-taxi conditions of other passengers and taxis in the domain can be masked while learning an RL policy. 

To summarize, the abstraction reasoner performs two specific steps -- first, it uses the domain knowledge as first-order logic statements to construct an abstract MDP and then grounds the MDP to construct the smaller ground MDP that an underlying Deep RL agent can solve. Automating these two steps fully, either using advances inside the lifted inference community or an LLM-based approach, remains an interesting future direction. At the lowest level are different RL agents, which, after acquiring the MDP, proceed to solve them appropriately. 


\begin{table*}[ht]
    \centering
    \footnotesize
   \begin{tabular}{|p{3cm}|p{7cm}|l|l|}
    \hline
    
    \textbf{\normalsize Domain} & \textbf{\normalsize D-FOCI statements} & \textbf{\normalsize Operators} & \textbf{\normalsize Relevant Predicates} \\
    \hline
    
    \normalsize Multiagent Taxi & 
    \parbox{10cm}{ 
        $\text{{taxi-at}}(T, L_1), \text{{move}}(\text{{T, Dir}}) \overset{+1}{\rightarrow} \text{{taxi-at}}(T, L_2)$ \\  
        $\text{{taxi-at}}(T, L), \text{{move}} (T, Dir)\overset{+1} {\rightarrow} \text{{R}} $\\
        $\text{{taxi-at}}(T1, L1), \text{{taxi-at}}(T2, L2), \text{{move}}(T1, Dir1),  $\\
        $\text{move}(T2, Dir2 )\overset{+1} {\rightarrow} \text{{R}} $ \\ \\
        $\mathbf{pickup(P, T):}$\\
        $\text{{taxi-at}}(T1,L1), \text{at}(P, L), \text{in-taxi}(P) \overset{+1} {\rightarrow} \text{in-taxi}(P) $\\ 
        $\text{in-taxi}(P) {\rightarrow} \text{{R}} $\\ \\
        $\mathbf{drop(P, T):}$\\
        $\text{{taxi-at}}(T1,L1), \text{in-taxi}(P), \text{dest}(P,L), \text{at-dest}(P)\overset{+1} {\rightarrow} \text{at-dest}(P) $\\ 
        $\text{at-dest}(P) {\rightarrow} \text{{R}} $ \\ 
    } & 
    \makecell{pickup(P, T) \\\\\\\\\\\\\ drop(P, T)} & 
    \makecell{\\ taxi-at$(T, L1)$, taxi-at$(T2,L2)$, \\ at$(P, L)$, in-taxi$(P)$, \\ move$(T,Dir)$, move$(T2,Dir2)$ \\ \\ \\ \\   taxi-at$(T, L1)$, taxi-at$(T2,L2)$,\\ dest$(P, L)$, in-taxi$(P)$, \\ move$(T,Dir)$, move$(T2,Dir2)$ }\\ 
    \hline
    
    \normalsize Multiagent Office & 
    \parbox{10cm}{ 
        $\text{{agent-at}}(A, L_1), \text{{move}}(\text{{A, Dir}}) \overset{+1}{\rightarrow} \text{{agent-at}}(A, L_2)$ \\  
        $\text{{agent-at}}(A, L_1), \text{{move}}(\text{{A, Dir}})  {\rightarrow} \text{{R}} $\\ \\ 
        $\mathbf{visitOrPickup(X, A):}$\\
        $\text{{agent-at}}(A,L1), \text{at}(X, L), \text{{with-agent}}(A,X) \overset{+1} {\rightarrow} \text{{with-agent}}(A,X) $\\ 
        $\text{{with-agent}}(A,X){\rightarrow} \text{{R}} $\\ \\
        $\mathbf{deliver(X, A):}$ \\ 
        $\text{{agent-at}}(A,L1), \text{{with-agent}}(A, X), \text{{office}}(L), $\\ 
        $\text{{delivered}}(X) \overset{+1} {\rightarrow} \text{{delivered}} $\\ 
        $\text{{delivered}}(X) {\rightarrow} \text{{R}} $ \\
    } & 
    \makecell{visitOrPickup(X, A) \\ \\\\\\\\deliver(X, A)} & 
    \makecell{$\text{{agent-at}}(A, L_1), \text{{move}}(\text{{A, Dir}}),$ \\
     $\text{at}(X, L), \text{{with-agent}}(A,X) $\\ \\ \\ \\ 
     $\text{{agent-at}}(A, L_1), \text{{move}}(\text{{A, Dir}}), $\\
     $\text{{with-agent}}(A,X), \text{{office}}(L),$ \\
     $\text{{delivered}}(X)$} \\
    \hline
    
    \normalsize Multiagent Dungeon & 
    \parbox{10cm}{ 
        $\text{{player-at}}(P, L1), \text{{orientation}}(P),\text{{move}}(P, Dir) \overset{+1}{\rightarrow} \text{{player-at}} (P,L2)$ \\ 
        $\text{{player-at}}(P, L), \text{{orientation}}(P), \text{{move}}(P, Dir){\rightarrow} \text{{R}} $ \\ \\         
        $\mathbf{attackEnemy (P, E):}$  \\
        $\text{{player-at}}(P,L1), \text{{enemy-at}}(E,L2), \text{{player-health}}(P), $\\
        $\text{{player-defense}}(P)  \overset{+1} {\rightarrow} \text{{player-health}}(P) $ \\ 
        $\text{{player-health}}(P) {\rightarrow}  R $\\ 
        $\text{{player-at}}(P,L1), \text{{enemy-at}}(E,L2), $ \\
        $ \text{{enemy-health}}(E) \overset{+1} {\rightarrow} \text{{enemy-health}}(E)$\\
        $\text{{enemy-health}}(E) {\rightarrow}  R $\\ \\
        $\mathbf{getKeyInDoor  (P, K)}: $  \\
        $\text{{hasKey}}(P, K), \text{{player-at}}(P, L1), \text{{enemy-at}}(E, L2) \overset{+1} {\rightarrow} \text{{hasKey}}(P, K) $ \\        
        $\text{{hasKey}}(P, K), \text{{doorUnlocked}}(K) \overset{+1} {\rightarrow} \text{{doorUnlocked}}(K) $ \\
        $\text{{doorUnlocked}}(K) {\rightarrow} \text{{R}} $ \\}  & 
        
    \makecell{attackEnemy(P, E) \\ \\ \\ \\ \\  \\ \\ getKeyInDoor(P, K) } & 
    \makecell{ $\text{{player-at}}(P, L1), \text{{orientation}}(P),$\\
    $ \text{{move}}(P, Dir), \text{{enemy-at}}(E,L2),$ \\
    $ \text{{player-health}}(P), \text{{player-defense}}(P), $\\
    $ \text{{enemy-health}}(E)  $ \\ \\ \\ \\
    $ \text{{player-at}}(P, L1), \text{{orientation}}(P),$ \\
    $ \text{{move}}(P, Dir), \text{{hasKey}}(P, K), $ \\
    $ \text{{enemy-at}}(E, L2), \text{{DoorUnlocked}}(K) $\\ 
    } \\
    \hline
    
\end{tabular}
    \caption{Example of D-FOCI statements for relational multiagent domains}
    \label{tab:example}
\end{table*}

\subsection{MaRePReL Algorithm}
Algorithm \ref{alg:MaRePReL} presents the procedure where we initialize the RL policies and buffers for various operators (\textbf{line 1}). The policies learned through our approach are a collection of task-specific operations. While one or more agents still have pending subtasks, we continuously collect trajectories from the environment for different operators, storing them in respective operator buffers (\textbf{lines 7-15}). In each episode, we first get the starting state from the environment \textbf{(line 4)}, and obtain the current tasks for each agent by computing their sub-plans using our relational multiagent planner, which is implemented by combining a SHOP \citep{shop} planner with branch and bound scheduling \textbf{(line 5-6)}. While one or more agents have some task remaining, we first compute the joint actions for the agents based on the current state, agent tasks, task policies, and the D-FOCI statements using the \textbf{GetAgentActions} method (\textbf{line 8}). Upon obtaining the joint action, we perform a step update using the \textbf{RePReLStep} method \textbf{(line 9)}. This step involves updating the state, buffers, plan, and tasks based on the D-FOCI rules for abstractions. Following RePReL's approach, the method returns the updated components along with a flag indicating the validity of the current plan. If the plan is considered invalid — one or more agents cannot perform the tasks assigned to them — then new agent-specific plans (subplans) are computed, and the agents are assigned new tasks \textbf{(lines 10-13)}. Once the episode has ended, we train a policy for each operator $\pi_o$ using a batch sampled from the buffer for the operator \textbf{(line 16-19)}. Once trained, the final policies for the different operators are returned. The methods \textbf{GetAgentActions} and \textbf{RePReLStep} are further detailed in the supplementary 
\footnote{Link to code and supplementary text: \url{https://starling.utdallas.edu/papers/MaRePReL}}
\section{Experimental Results}

\label{sec:results}

We present our results across different tasks in three relational multiagent domains that demonstrate the effectiveness of MaRePReL. We answer the following questions explicitly. 
\begin{enumerate}[wide,  labelindent=0pt]
\itemsep0em 
    \item Does \textbf{MaRePReL} improve {\bf sample efficiency} ?
    \item Does \textbf{MaRePReL} efficiently {\bf transfer} from one task to another?
    \item  Does \textbf{MaRePReL generalize} to varying number of objects?
\end{enumerate}
\subsection{Domains}
For the first environment, we extend the taxi domain environment \citep{Diettrich2000} to relational multiagent settings. The goal is to transport passengers from their current locations to their destinations. Passengers are located at four different grid positions -- R, G, B, and Y -— requiring coordinated efforts from the taxis for pickup and drop-off, with no two passengers having the same pickup or drop locations. Additionally, the taxis cannot cross each other or occupy the same location. Doing so would cause crashes, terminating a huge negative reward and incurring a heavy penalty.

For the second environment, we extend the office world domain \citep{icarte2018rm} to accommodate multiple agents. The agents are presented with a set of tasks they need to complete together. A positive reward is provided to the agents when they complete the tasks, but a bump penalty is given to the agents when they move into the same cell. 

For the third environment, we've handcrafted a dungeon grid world inspired by Unity's Dungeon Escape \citep{unityml}. The agents must defeat different enemies, collect the keys required to unlock the door, and escape the dungeon. The enemies, if attacked, will target the agents back. The agents become incapacitated and unable to take action once their health reaches zero.  The environment terminates once all the keys are in the door, or all the agents are dead.

Table \ref{tab:example} lists the D-FOCI statements for the three environments with additional environment details in the supplementary. 
\renewcommand{\arraystretch}{1}
\begin{table*}[ht]
    \centering
    \begin{tabular}{|c|c|c|c|}
    \hline
        \textbf{Environment} &  \textbf{Task 1} &\textbf{ Task 2} & \textbf{Task 3} \\\hline
         Multiagent Taxi& Transport 2 passengers &  Transport 3 passengers  & Transport 4 passengers \\\hline
         Multiagent Office & Visiting A, B, C and D & Pickup mail and coffee & Deliver mail and coffee to office \\\hline
         Multiagent Dungeon & Defeat 1 skeleton and escape & Defeat 1 dragon and escape & Defeat 1 skeleton, 1 dragon and escape\\\hline
    \end{tabular}
    \caption{Tasks to perform in different environments}
    \label{tab:tasks}
\end{table*}

\subsection{Baselines}
\begin{figure*}[!ht]
    \centering
      \begin{subfigure}{0.32\linewidth}
        \includegraphics[width=\linewidth]{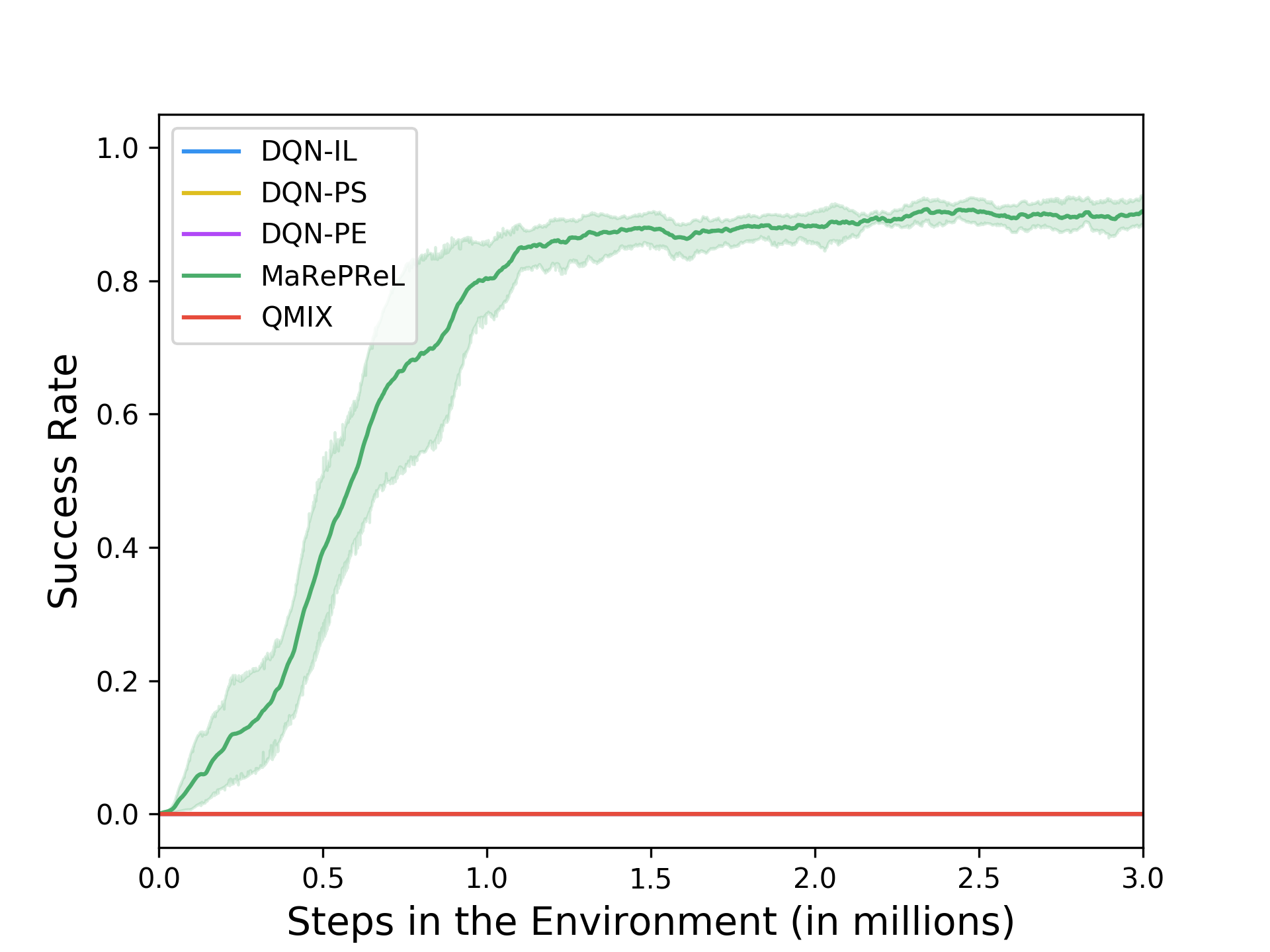}
        \caption{Taxi Task 1}
    \end{subfigure}
      \begin{subfigure}{0.32\linewidth}
        \includegraphics[width=\linewidth]{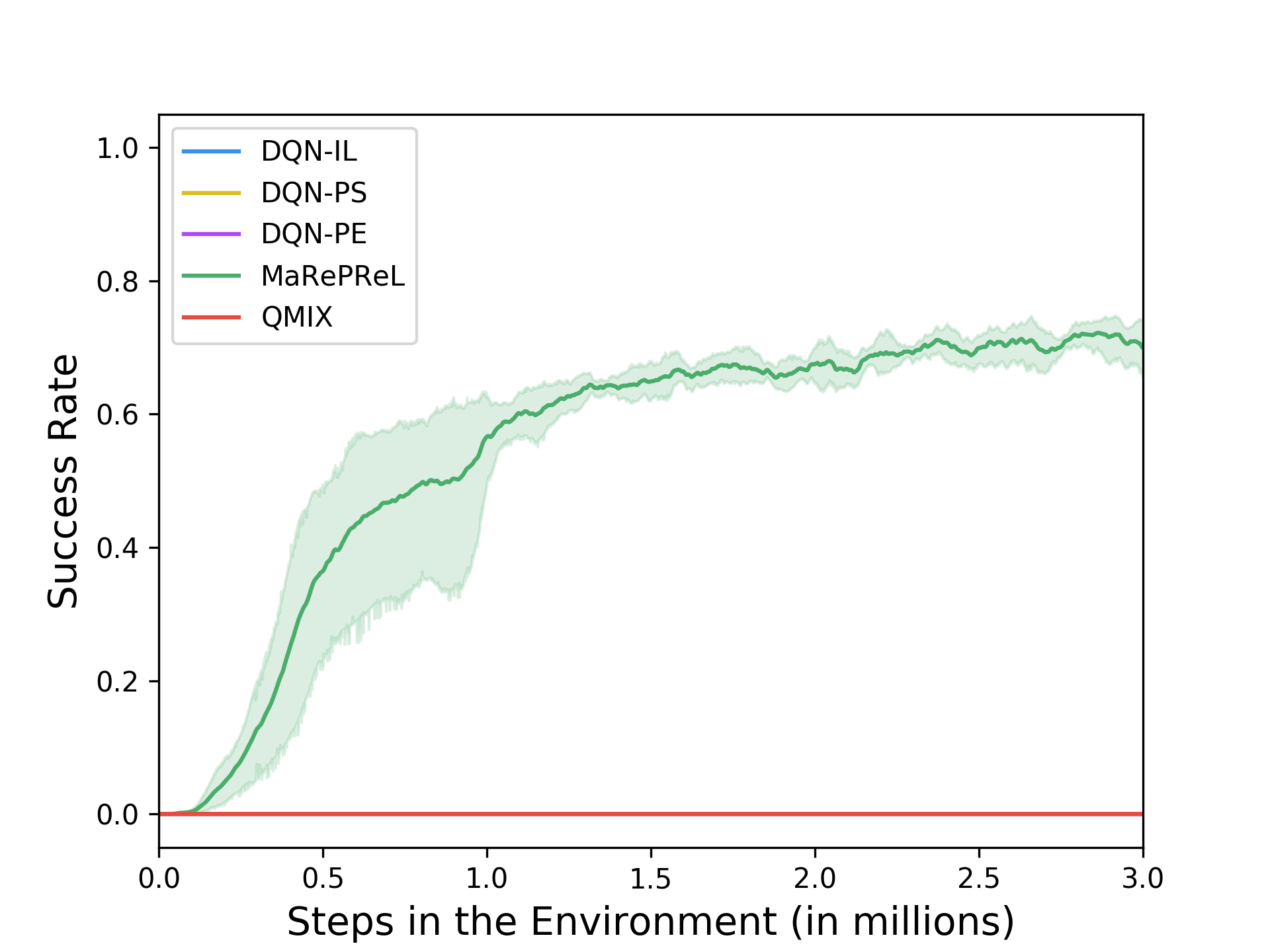}
        \caption{Taxi Task 2}
    \end{subfigure}
      \begin{subfigure}{0.32\linewidth}
        \includegraphics[width=\linewidth]{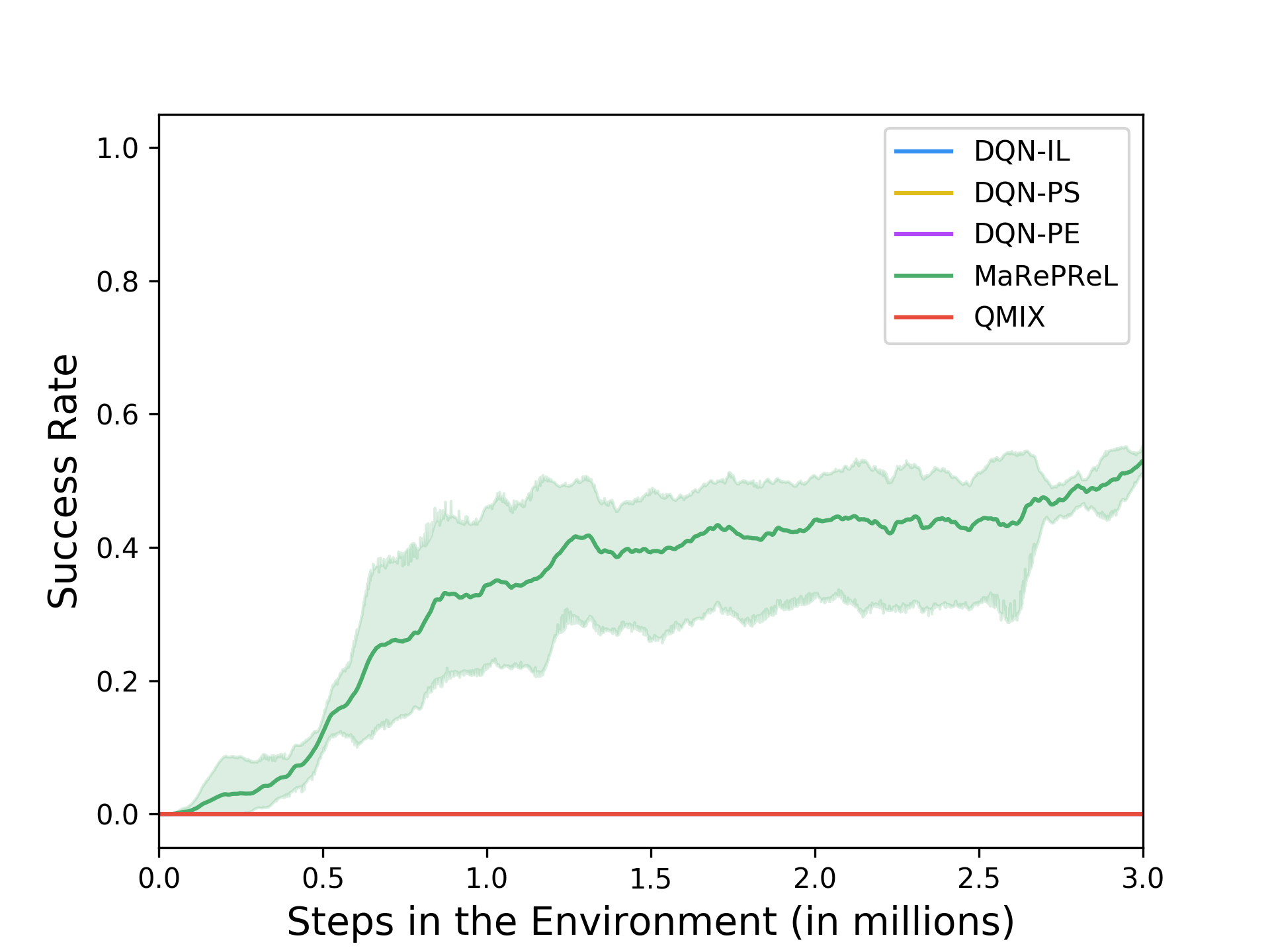}
        \caption{Taxi Task 3}
    \end{subfigure}
        \centering
    \begin{subfigure}{0.32\linewidth}
        \includegraphics[width=\linewidth]{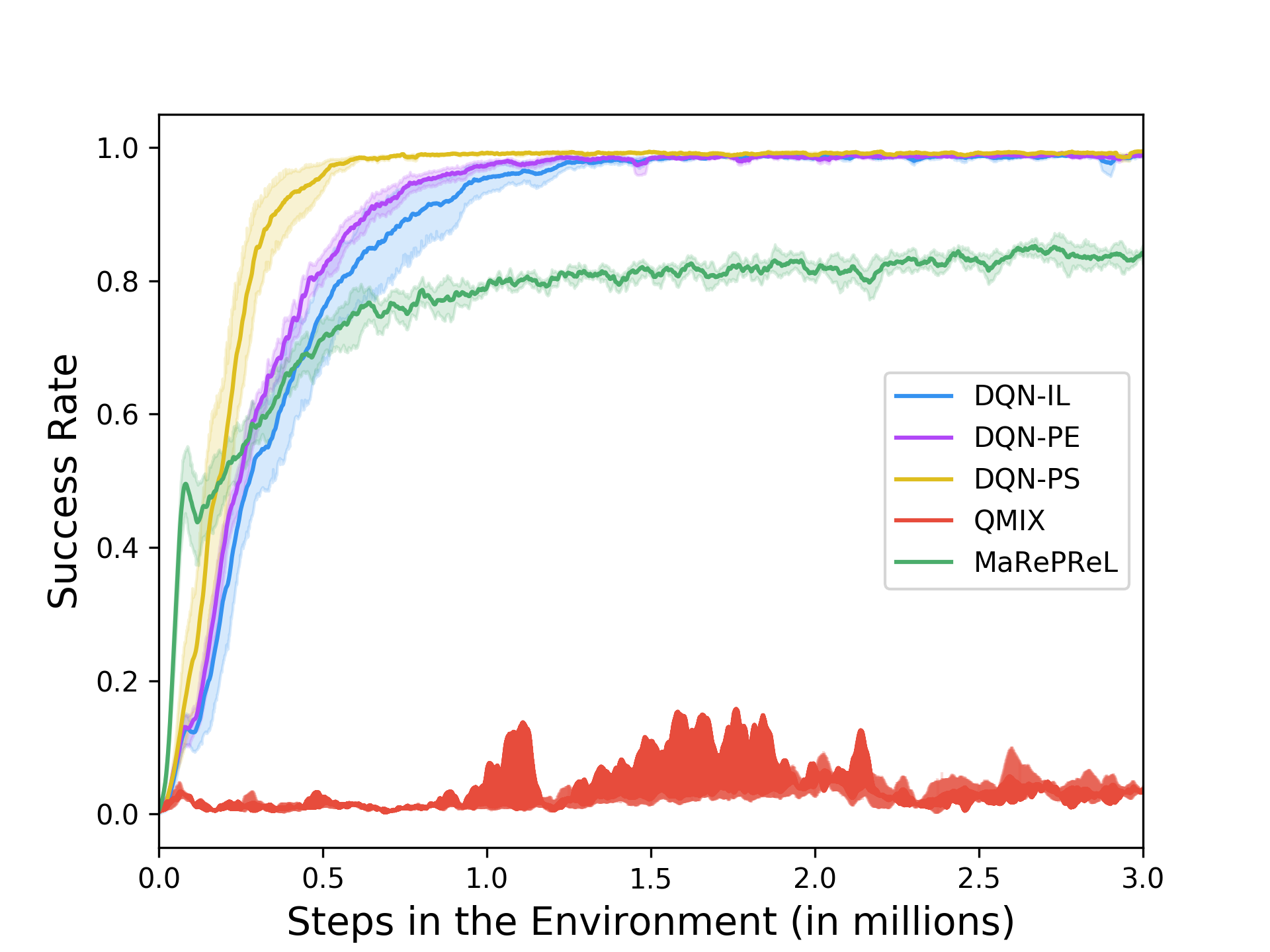}
        \caption{Office Task 1}
    \end{subfigure}
    \begin{subfigure}{0.32\linewidth}
        \includegraphics[width=\linewidth]{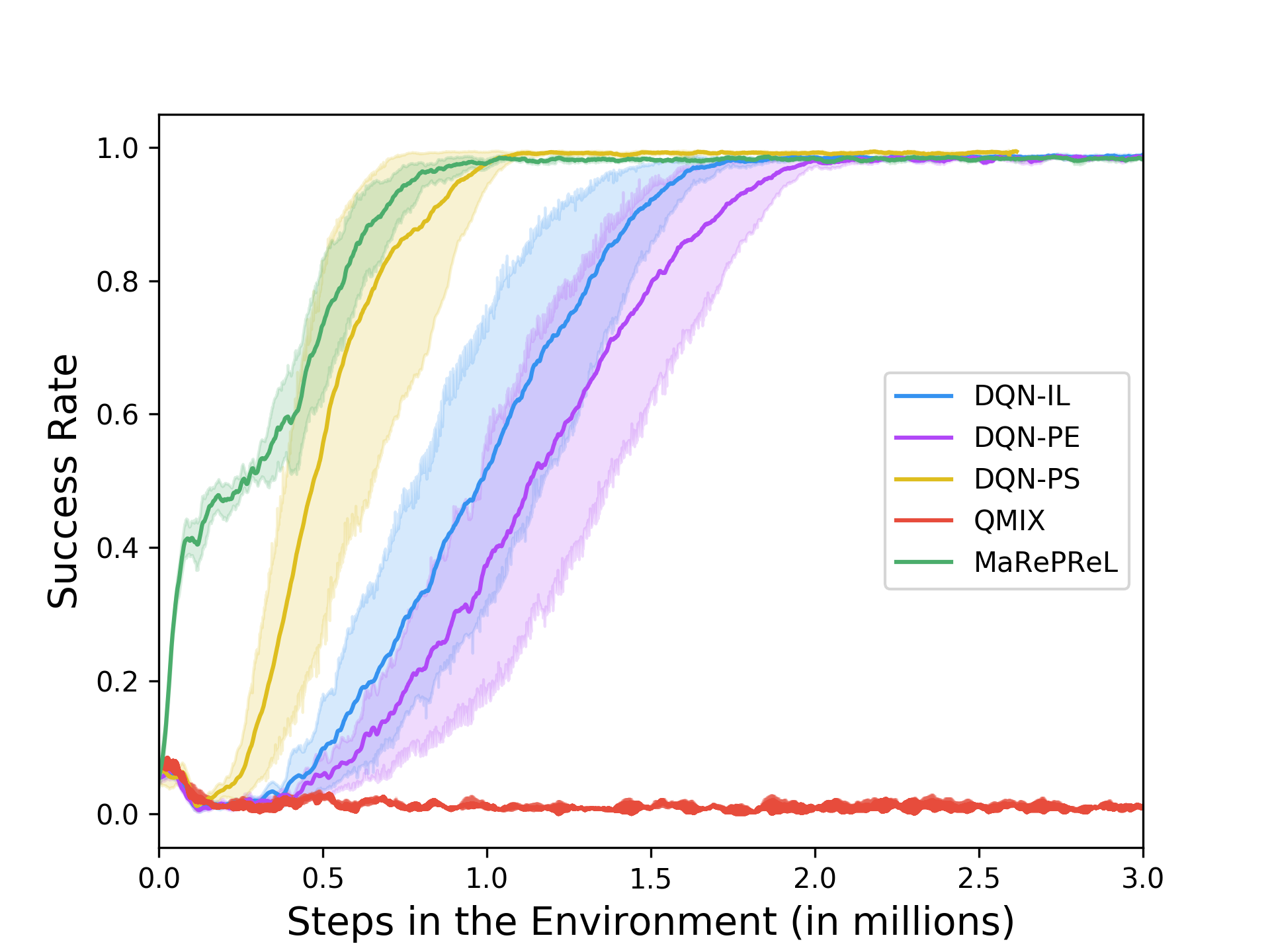}
        \caption{Office Task 2}
    \end{subfigure}
    \begin{subfigure}{0.32\linewidth}
        \includegraphics[width=\linewidth]{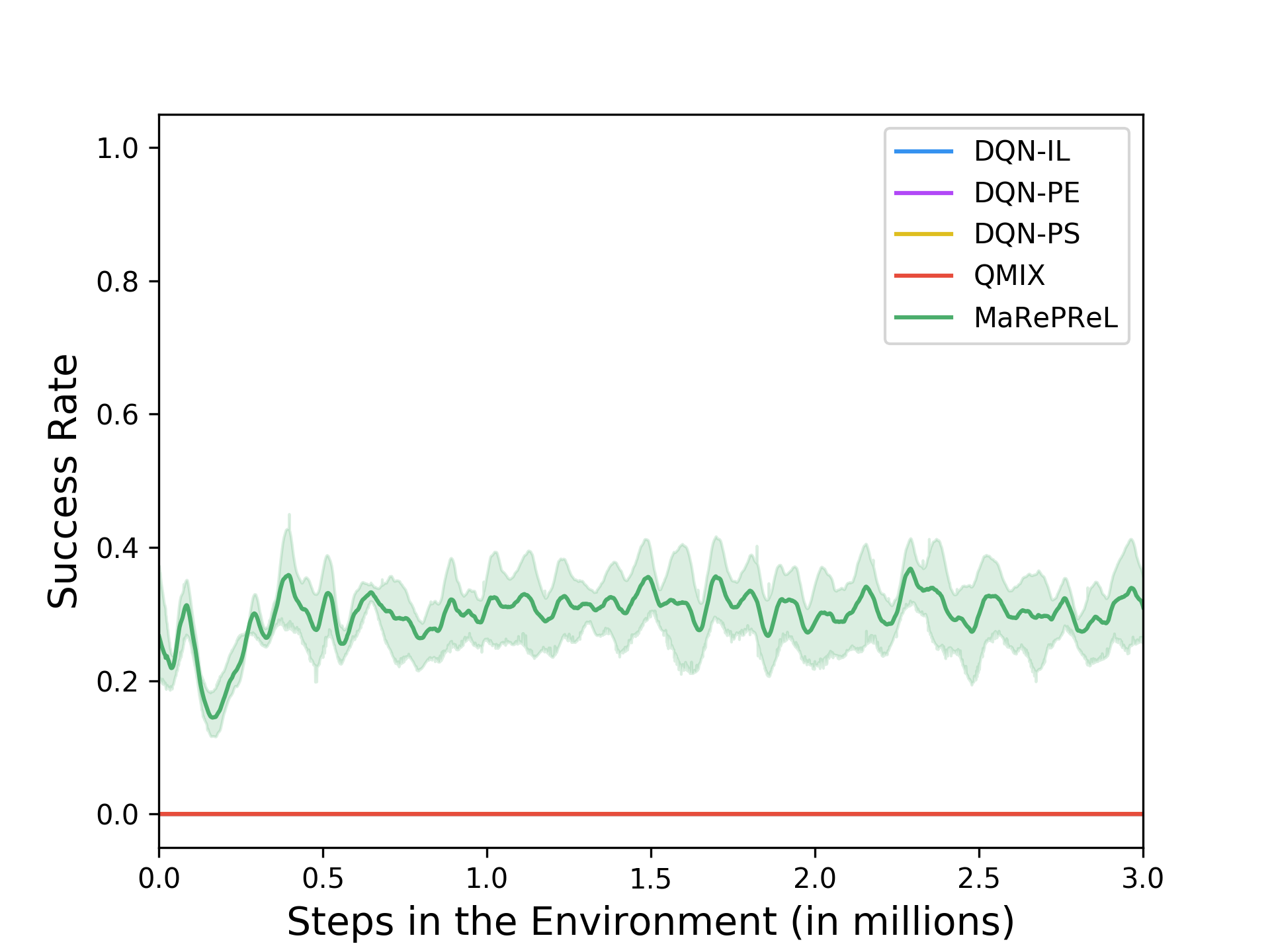}
        \caption{Office Task 3}
    \end{subfigure}
    \centering
      \begin{subfigure}{0.32\linewidth}
        \includegraphics[width=\linewidth]{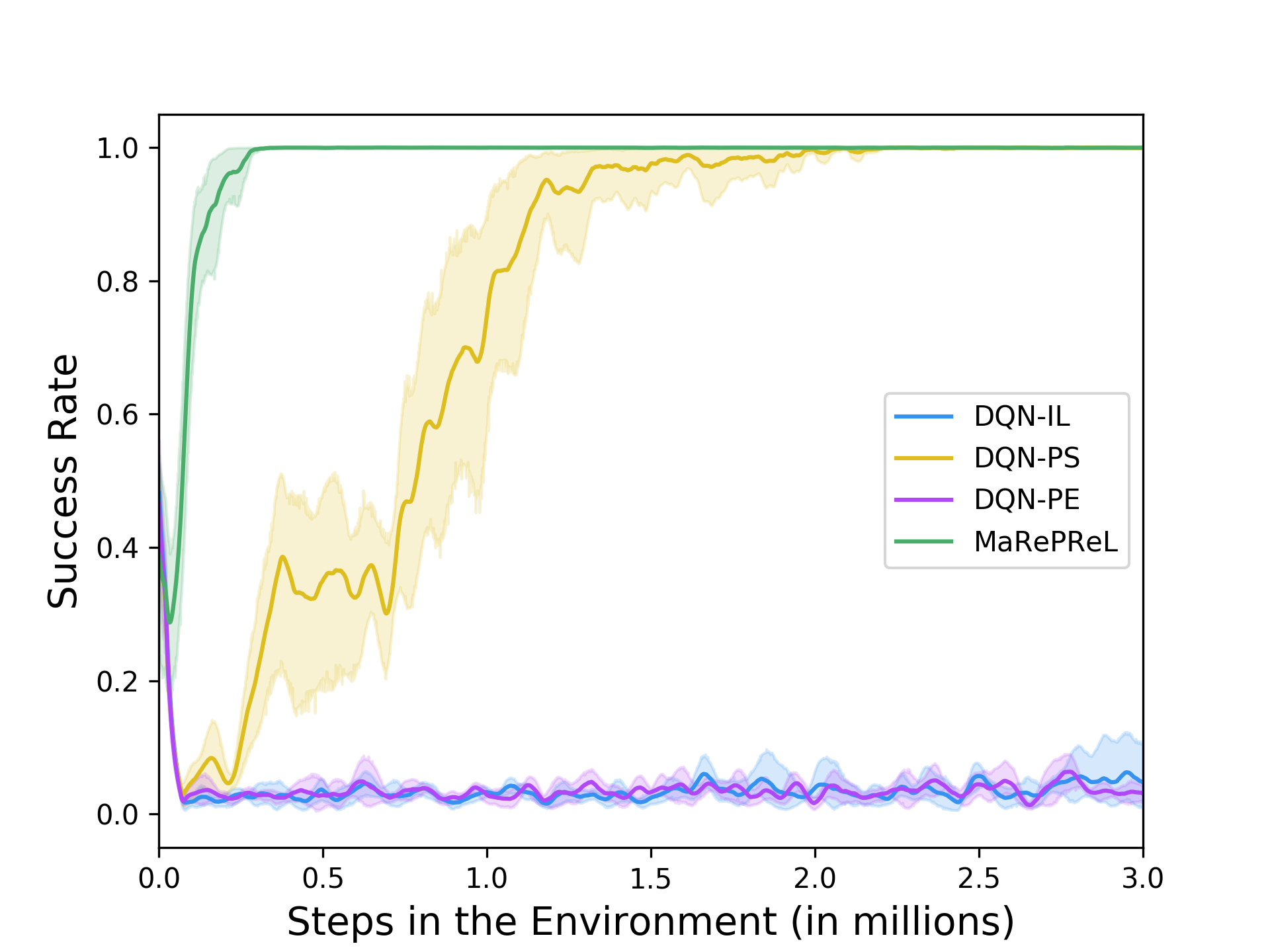}
        \caption{Dungeon Task 1 }
    \end{subfigure}
      \begin{subfigure}{0.32\linewidth}
        \includegraphics[width=\linewidth]{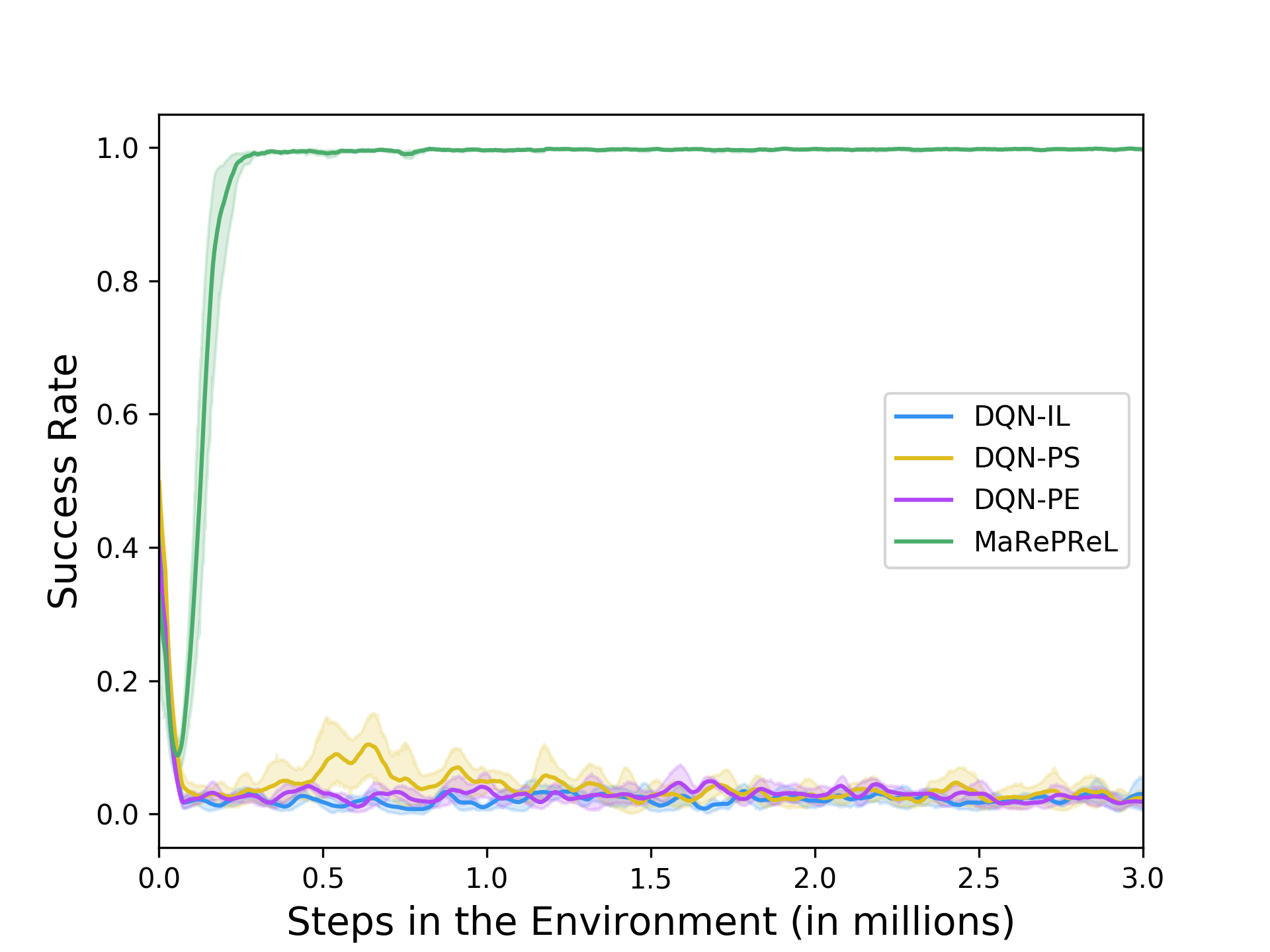}
        \caption{Dungeon Task 2 }
    \end{subfigure}
      \begin{subfigure}{0.32\linewidth}
        \includegraphics[width=\linewidth]{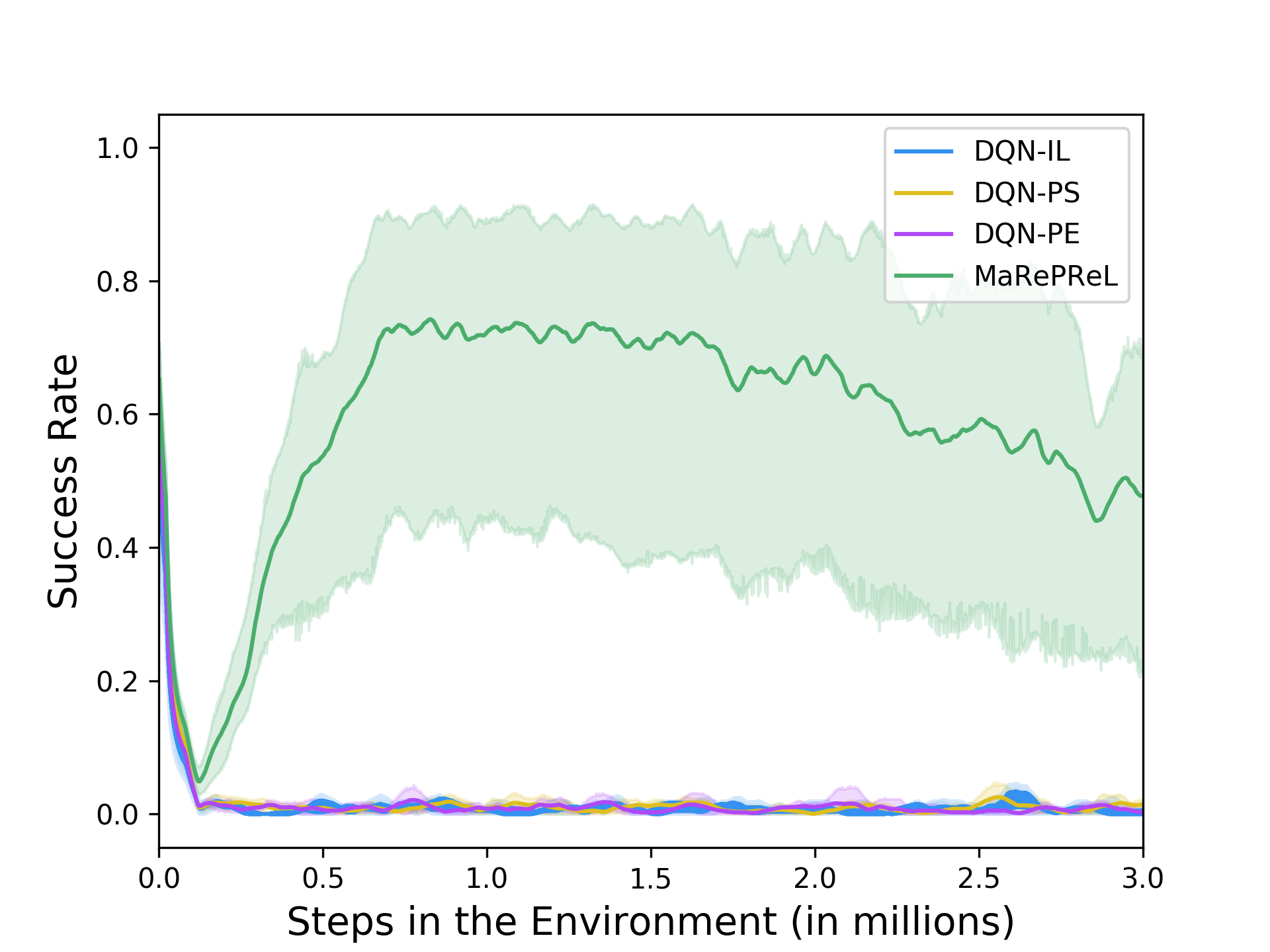}
        \caption{Dungeon Task 3}
    \end{subfigure}
    \caption{The success rates across different methodologies for the Taxi, Office World, and Dungeon domains }
    \label{fig:success_rate}
    \Description[Description of the results on the 9 tasks across 3 domains]{MaRePReL outperforms the baselines everywhere except the tasks in the office world due to the baselines being able to memorize the simple domain}
\end{figure*}
We evaluate MaRePReL against several standard MARL algorithms, including Deep Q-Networks with parameter sharing (DQN-PS), Deep Q-Networks as independent learners (DQN-IL), and QMIX \citep{Qmix}. In DQN, each agent maintains its decentralized state-action value function, updating Q-values based solely on local observations and individual rewards. In contrast, QMIX utilizes a parameterized mixing network to compute a joint Q-value, combining information from all agents. However, it is essential to note that QMIX is not included in the benchmark for the multi-agent dungeon environment due to a key limitation: QMIX requires a fixed number of agents to function properly, which is not guaranteed in the dungeon environment as the number of agents change when an agent dies.

Additionally, we introduce a new baseline called DQN-PE (DQN with Plan Embeddings), inspired by HTN-MTRL \citep{htn_mtrl}. In DQN-PE, the observation for each agent is augmented with a vector embedding of its current sub-plan. This approach leverages the same HTN planner as MaRePReL and encodes the sub-plan string using a BERT embedding model \citep{sbert}. The string is divided into chunks, mean pooled, and reduced to a $\mathbb{R}^4$ embedding. This baseline allows us to assess whether providing agents with task-specific information in the form of the agent's sub-plan embeddings is sufficient, or whether task hierarchies and state abstractions are required like in MaRePReL.  As far as we know, no relational multiagent baselines are readily available. One could design a relational multiagent baseline by considering hierarchies as a special form of relations (in the lines of the work inside RRL community~\citep{FODDs, DasFittedQ}), but extending them to a multiagent scenario is non-trivial and outside the scope.  

\begin{figure*}
    \centering
      \begin{subfigure}{0.32\linewidth}
        \includegraphics[width=\linewidth]{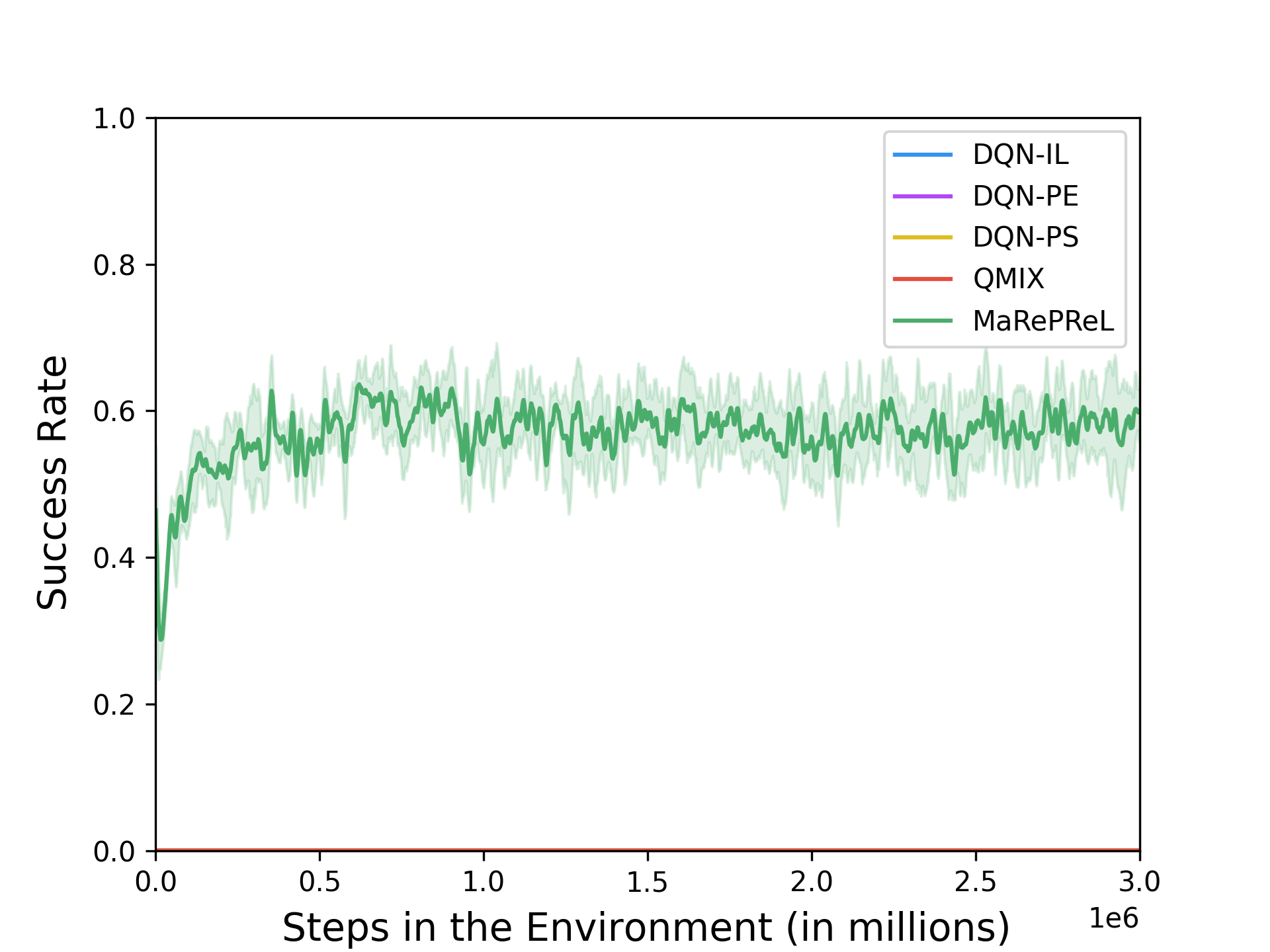}
        \caption{Taxi Task 1 to Task 3}
    \end{subfigure}
    \begin{subfigure}{0.32\linewidth}
        \includegraphics[width=\linewidth]{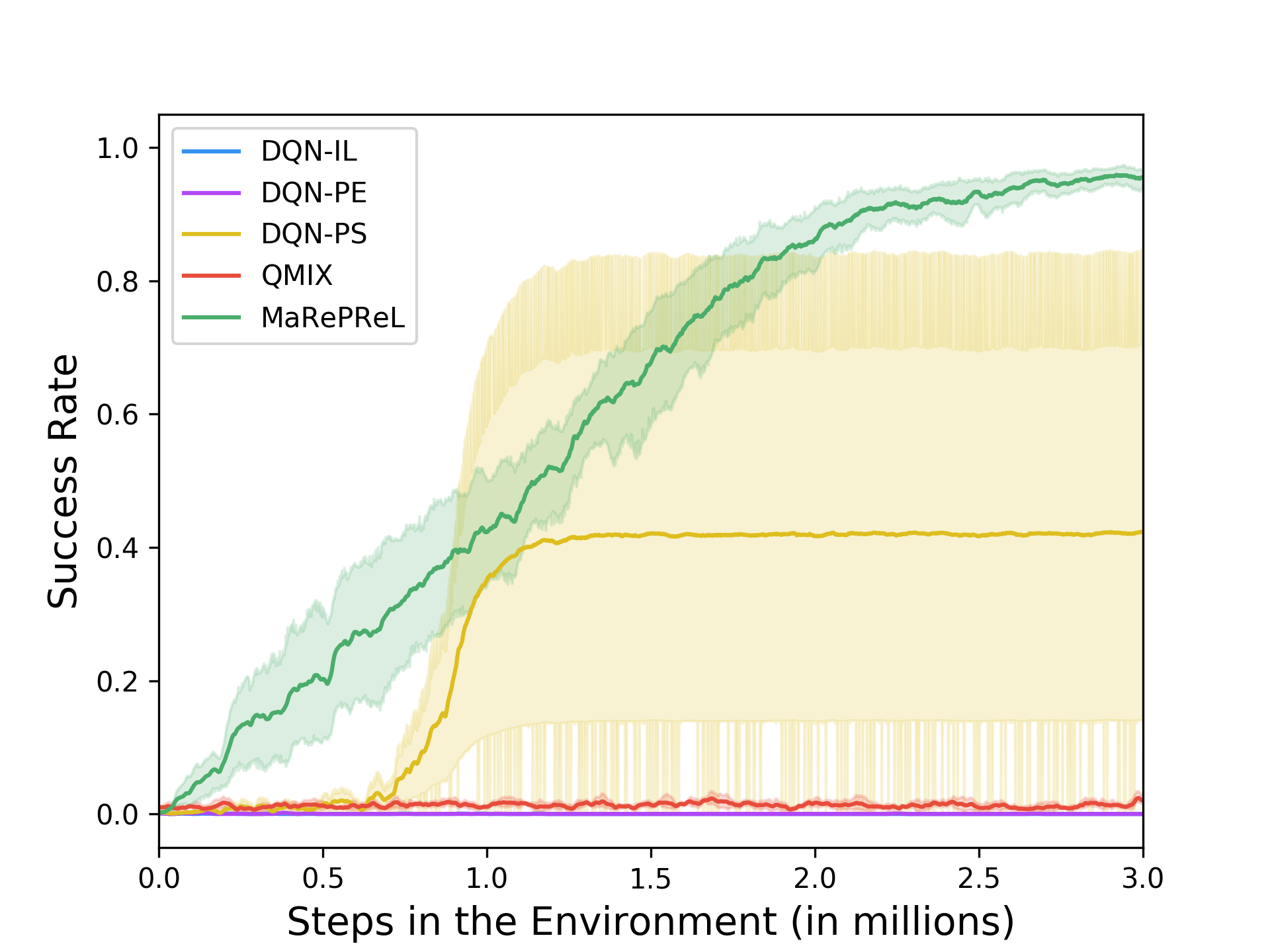}
           \caption{Office Task 1 to Task 2}
    \end{subfigure}
    \begin{subfigure}{0.32\linewidth}
        \includegraphics[width=\linewidth]{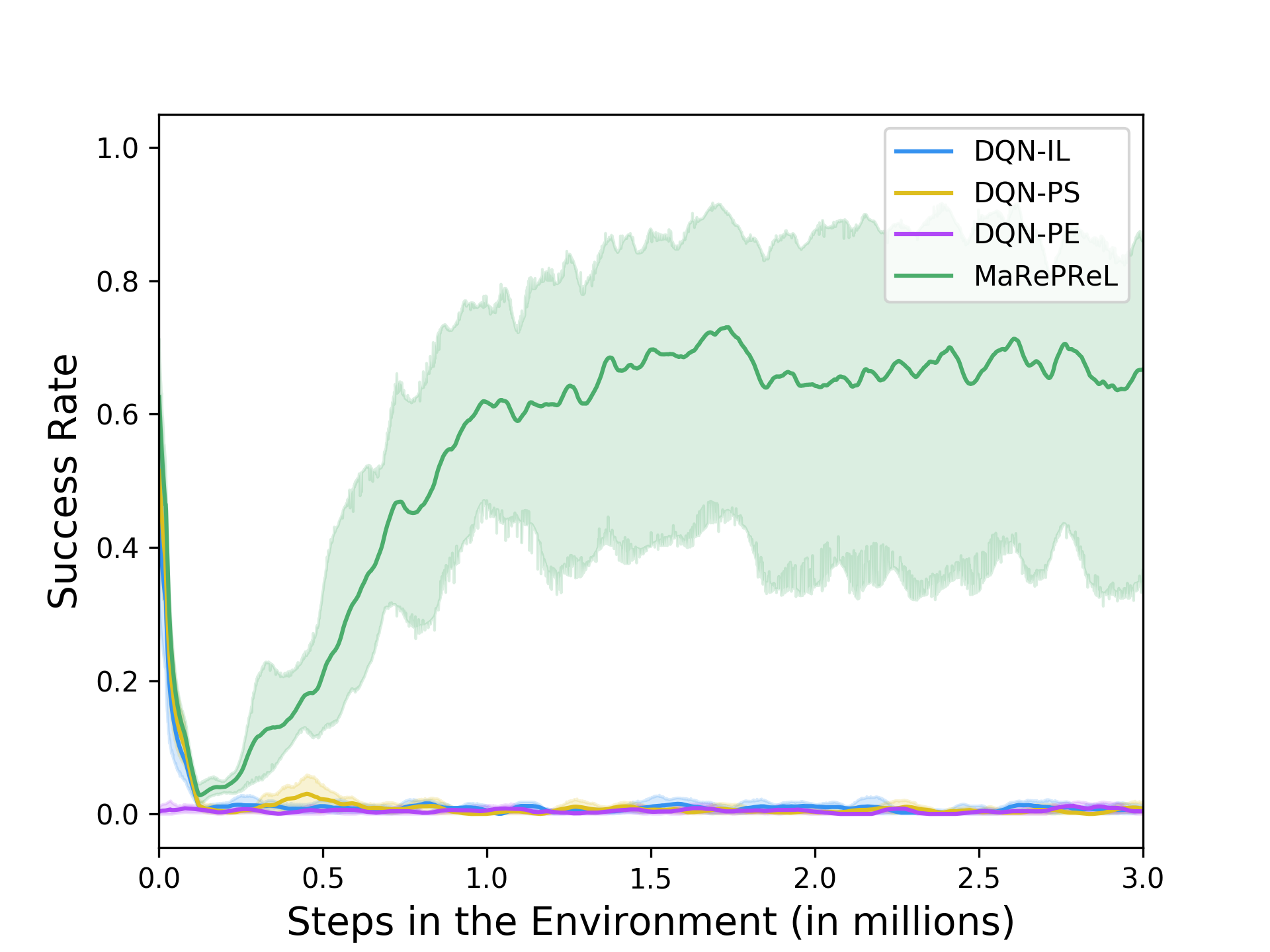}
        \caption{Dungeon Task 2 to Task 3}
    \end{subfigure}
    \caption{The success rate in case of transferring from policy for one task to another across different methodologies for the Taxi, Office World, and Dungeon Domains}
    \label{fig:transfer_gen}
    \Description[The transfer and generalization performance of the model]{The transfer and generalization performance of the model across different tasks, MaRePReL seems to be outperforming the baselines}
\end{figure*}
\subsection{Results}
We evaluate sample efficiency, transfer ability and generalization capability of our method against the other baselines across the different tasks (Table \ref{tab:tasks}). The results, averaged over five trials, are presented in Figure \ref{fig:success_rate} (training from scratch), and \ref{fig:transfer_gen} (training starting with a previously learned policy), where the bold line represents the mean and the shaded region illustrates the variance of the success rate across trials after 3 million environment steps.

\paragraph{Sample Efficiency:} In the taxi domain, MaRePReL, unlike DQN and QMIX, was able to learn how to complete tasks 1, 2, and 3, whereas DQN-IL, DQN-PS, DQN-PE and QMIX have a near-zero success rate even after training for 3 million steps (Figure \ref{fig:success_rate} a-c). In the Office World domain, DQN-PS learned an optimal policy for Tasks 1 and 2 (Figure \ref{fig:success_rate} d-e). This can be attributed to the static nature of the goals in the environments. However, as the complexity of the tasks increased, such as in Task 3 (Figure \ref{fig:success_rate} f), none of the baselines could perform while MaRePReL still was able to show early success. In the Dungeon domain, MaRePReL demonstrates robust performance, converging to optimal solutions for Task 1 and 2 while showing a steeper learning curve for Task 3. The baseline of DQN-PS exhibits convergence for Task 1 after one million steps, although notably slower compared to MaRePReL. However, DQN-IL, DQN-PS, and DQN-PE struggled with learning Tasks 2 and 3, showing a near-zero success rate in both cases (Figure \ref{fig:success_rate} g-i). QMIX can't be used in the dungeon environment due to its changing number of agents. Therefore, \textbf{Q1} can be answered affirmatively.

\paragraph{Transfer:} 
In the Office World domain, we transfer policies from Task 1 (Visiting A, B, C, D) to Task 2 (Get Mail and Coffee), and in the Dungeon domain, from Task 2 (Dungeon with 1 Dragon) to Task 3 (Dungeon with 1 Dragon and 1 Skeleton). In the Office World experiment, MaRePReL successfully adapted to the new task and nearly achieved 100\% success, whereas the baselines struggled— QMIX, DQN-IL, and DQN-PE failed, and DQN-PE had a low success rate (Figure \ref{fig:transfer_gen}b). In the Dungeon experiment, when tasks shared common goals, transferring with MaRePReL facilitated convergence, while neither QMIX nor any DQN baselines showed success (Figure \ref{fig:transfer_gen}c). These results demonstrate the transfer abilities of a relational model, a fact well-known within the relational RL community. The key power of relational models (in our case, relational planner and relational abstract reasoner) lies in their ability to achieve efficient transfer and effective generalization. Therefore \textbf{Q2} can be answered affirmatively.

\paragraph{Generalization:}
In this case, the policies are not randomly initialized; instead, the policies learned for tasks with fewer objects are applied to the new task with more objects. For the taxi world domain, a policy trained on the task of transporting two passengers is applied to the task of transporting four passengers (Figure \ref{fig:transfer_gen}a). MaRePReL significantly improves sample efficiency, achieving Task 3's success rate in less than half a million steps, compared to 3 million steps for the non-transferred policy. Similarly, one can notice that MaRePReL generalizes to a task presenting an increasing number of enemies to defeat in the Dungeon environment. Other baselines do not demonstrate any initial success or show any performance improvement. Therefore, \textbf{Q3} can be answered affirmatively
\balance
\section{Discussion and Future Work}
\label{sec:discussion}
We demonstrated empirically that MaRePReL significantly outperforms traditional MARL approaches, including DQN (independent learners or parameter sharing or with sub-plan embeddings) and QMIX. Our results clearly show the effectiveness of combining a relational planner with an agent-specific task distributor at the higher level and deep RL at the lower level. Significant improvements can be observed in both learning, transfer, and generalization.

Our framework has a few limitations. As the number of operators and agents increases, the search space for the relational planner grows exponentially, posing challenges for generalization, an important future direction. Our current formalism applies only to problems featuring a fully observable state space, and extending it to partially observable spaces necessitates integration with efficient (lifted) probabilistic inference. Moreover, the cooperation shown between agents is loosely coupled as they work in parallel to complete the tasks assigned by a centralized planner. It is possible to extend our approach to tackle challenges in domains that demand coordination among multiple agents \citep{samvelyan2019starcraft2, christianos2020rware} by incorporating a partial-order planner along with wait operators. This extension would allow all agents to achieve a state that fulfills the preconditions before performing the joint task. Finally, constructing a fully differentiable system is an interesting direction for future research.



\begin{acks}
NP, RS, and SN gratefully acknowledge the support of the ARO award W911NF2010224. PT gratefully acknowledges the support of ARO award  W911NF2210251.
\end{acks}

\bibliographystyle{ACM-Reference-Format} 
\bibliography{bibliography}

\clearpage
\onecolumn
\appendix

\begin{center}
    \Huge \textbf{Supplementary}
\end{center}

\section{RePReL}
RePReL \cite{kokel2021reprel} is a hierarchical framework that integrates relational planning and reinforcement learning to solve goal-directed sequential decision-making problems. It leverages the strengths of both approaches to improve convergence and enable effective transfer across multiple tasks. The framework uses a high-level relational planner to decompose goals into subgoals. These subgoals are then passed to a low-level reinforcement learning agent that tries to achieve them with minimal cost. RePReL  \footnote{ For more on RePReL: \url{https://starling.utdallas.edu/papers/RePReL/}}adapts first-order conditional influence (FOCI) statements to specify bisimilarity conditions of MDPs, justifying safe and effective abstractions for reinforcement learning. The relational representation used in its hierarchical ordered planning facilitates generalization across varying numbers and types of objects without requiring excessive feature engineering. Empirical evaluations demonstrate RePReL's superior performance, efficient learning, and generalization to unseen tasks compared to other planner-RL combinations, such as TaskableRL. RePReL's ability to generalize across a varying number of objects makes it well-suited for goal-directed relational domains. 

\section{Task Distributor}
The output of the planner is typically the task decomposition and does not bind the tasks to the specific agents. We use a task distributor as part of the relational planner to divide the ordered plan provided into agent-specific sub-plans using agent constraints for the different tasks. The grounded plan (high-level plan) and a set of ordering constraints are used to distribute the tasks to create agent-specific plans (sub-plans). A greedy approach is used to schedule tasks that involve forward chaining \citep{kvarnstrom2011planning}. It examines the causal links between operators and prevents tasks from being assigned to agents that cannot execute them. 

A \textbf{causal link} is defined as an edge between grounded operator $O_p$ and $O_q$ if there exists a literal in the effects of $O_p$ that is also part of the preconditions of $O_q$. The set of causal links is identified after a plan is derived. The causal links $L = (l_1, l_2, \cdots, l_n)$ define a partial ordering between operations. Our task distributor ensures that the same agent performs the causally linked operations. 

The current domains shown in our experiment section assume that a plan can be split into $N$ causally independent sub-plans, where $N$ can be the number of tasks or sub-tasks. As mentioned in Section 3.2, our current implementation of the distributor solely utilizes a greedy approach with causally linked operators to efficiently allocate sub-plans. Sub-tasks are distributed equally, with the current implementation treating each subtask with a cost of 1. The agent-specific task distributor focuses on maintaining an equal task execution cost between agents. Our framework can be easily extended to handle more complex cost functions as there is no inherent assumption about the costs. While we used homogeneous agents for experiments, there is no inherent assumption about the types of agents, and hence heterogeneous agents are easily implementable in the framework. The task distributor for heterogeneous agents would include a cost function that incorporates both task-specific information and the agent's properties. This detail can account for different agent skills for more complex versions of domains such as the dungeon environment, where we can assign the agents different classes that deal different damages based on enemy type.

\section{MaRePReL Methods}
We describe the two subroutines used by the MaRePReL algorithm, i.e. \textbf{GetAgentActions} and \textbf{RePReLStep} in detail

\subsection{GetAgentActions}
The method \ref{alg:getagentactions} is used to get the agent actions given the current state, the current tasks for each agent, the current policy, the D-FOCI statements, and the set of agents. 

We initialize each agent's actions dictionary with $NULL$ (no-action). We iterate over the different agents, getting the policy of the agents based don't the current task (which decides the operator) \textbf{(lines 2-12)}. Moreover, we use D-FOCI statements $F$ along with our current operator to construct the abstract state $\hat{s}$ from our current state $s$ \textbf{(line 6)} by masking out the irrelevant parts of the current state for the task at hand. 
We evaluate the current state against the terminal condition for the operator to determine whether it is terminal. In case the state is non-terminal, we use the policy to get a new action \textbf{(lines 7-10)}. Once we have determined the actions for all the agents, we return a dictionary of their actions \textbf{(line 13)}.
\begin{algorithm}[ht]
    \caption{GetAgentActions}
    \begin{algorithmic}[1]
        \Statex \textbf{Input:} Current state $s$, current agent tasks $\phi$, current operator policies $\pi_o$, the D-FOCI statements $F$, and the set of agent $A$
        \Statex \textbf{Output:} The dictionary of actions for the different agents $actions$
        \Statex
        \State Initialize the actions dictionary $actions$ with $NULL$ action for each agent
        \For{agent $i \in A$}
        \If{current task for agent $i, \phi_i$ is not $NULL$}
        \State Get the current operator $o_i$ based on current task $\phi_i$
        \State The agent policy, $\pi_i \gets \pi[o_i]$
        \State Get the abstract state $\hat{s}$ based on the current state $s$, 
        \Statex \quad \quad \quad operator $o_i$, and D-FOCI statements $F$
        \State $terminalState \gets \hat{s} \in \beta(o_i)$
        \If{$\neg terminalState$}
        \State The action for the agent $actions[i] \gets \pi_i(\hat{s})$
        \EndIf
        \EndIf
        \EndFor
        \State \textbf{returns} $actions$
    \end{algorithmic}
    \label{alg:getagentactions}
\end{algorithm}

\begin{algorithm}[ht]
\caption{RePReLStep}
    \begin{algorithmic}[1]
        \Statex \textbf{Input:} The current state $s$,  the dictionary of agent actions $actions$, the operator buffers $\mathcal{D}$,  the terminal reward $t_R$, the current task for the agents $\phi$, the current plan $\Pi$, the D-FOCI rules $F$, and the set of agents $A$
        \Statex \textbf{Output:} The next state s', the updated buffers $\mathcal{D}$, the updated tasks for the different agents $\phi$, and the plan flag $planValid$
        \Statex
        \State Perform the step in the environment with state $s$ using $actions$ to get the next state $s'$ and agent rewards $rewards$
        \State $planValid \gets True$
        \For{each agent $i \in A$}
        \If{the agent task $\phi_i$ is $\neg NULL$}
        \State Get the current operator for the agent  $i, o_i$ 
        \Statex \quad \quad \quad based on the current task $\phi_i$
        \State Get the abstract states $\hat{s}, \hat{s'}$ using the operator $o_i$ and 
        \Statex \quad \quad \quad D-FOCI statements $F$ for the states $s$ and $s'$ respectively
        \State $wasTerminal \gets \hat{s} \in \beta(o_i)$
        \State $preConditionMet \gets \hat{s} \in I(o_i)$
        \State $isTerminal \gets \hat{s'} \in \beta(o_i)$
        \If{$isTerminal$}
        \State $rewards[i] \gets rewards[i] + t_R $
        \State $\phi_i \gets $ Pop the first task from agent plan $\Pi_i$
        \Else
        \If{$wasTerminal$ or $\neg preConditionMet$}
        \State $planValid \gets False$
        \EndIf
        \EndIf
        \State $\mathcal{D}_{o_i} \gets \mathcal{D}_{o_i} \cup \{\hat{s}, actions[i], rewards[i], \hat{s'}\}$
        \EndIf
        \EndFor
        \State \textbf{return} $s' , \mathcal{D}, \phi, planValid$
    \end{algorithmic}
    \label{alg:performstep}
\end{algorithm}
\subsection{RePReLStep}
The method \ref{alg:performstep} given the current state $s$, actions for each agent $actions$, the operator buffers $\mathcal{D}$, the terminal reward $t_R$, the current tasks for the different agents $\phi$, the current plan $\Pi$, the set of D-FOCI statements $F$ and the set of agents $A$, would perform a step in the environment to return the next state $\hat{s}$, the updated trajectory buffers $\mathcal{D}$, agent tasks $\phi$ as well as the flag indicating whether the current plan is valid $planValid$.

The method begins with taking the joint step for the agents in the environment based on the current state $s$, and agent actions $actions$ to get the updated state $s'$ and agent rewards $rewards$ \textbf{(line 1)} and we initially assume that the current agent-specific plans are valid \textbf{(line 2)}. 

We iterate over each agent, and for agents with pending tasks, we get the current operator for the agent \textbf{(line 5)}, the abstract states $\hat{s}, \hat{s'}$ based on the D-FOCI rules $F$ and operator for the agent $o_i$ \textbf{(line 6)}. We compute whether the current state or the previous state was terminal \textbf{(line 7-8)}. 
Additionally, we use the multiagent planner (which defines all the operators) to the current state against the terminal condition of the agent's operator. If the current state is terminal, we add the terminal rewards to the agent's reward and update the agent task based on the plan $\Pi$ \textbf{(line 10-12)}. Additionally, we check whether the previous state was terminal and whether the previous state satisfies the precondition of the operator \textbf{(line 8-9)}. If the previous state was terminal, or the precondition was not satisfied, it means that our plan is no longer valid, and we would need to compute it again \textbf{(line 12-15)}. We add the current transition for the agent to the agent's operator buffer \textbf{(line 16)}. Once all the agents have been evaluated, we return the successor state $s'$, the updated buffer $\mathcal{D}$, the updated agent tasks $\phi$, and the $planValid$ flag \textbf{(line 21)}

\section{Experiment Domains}
Three relational multiagent domains were considered for the experiments, which are described in the subsequent subsections
\begin{figure*}[t]
    \centering
      \begin{subfigure}{0.28\linewidth}
        \includegraphics[width=\linewidth]{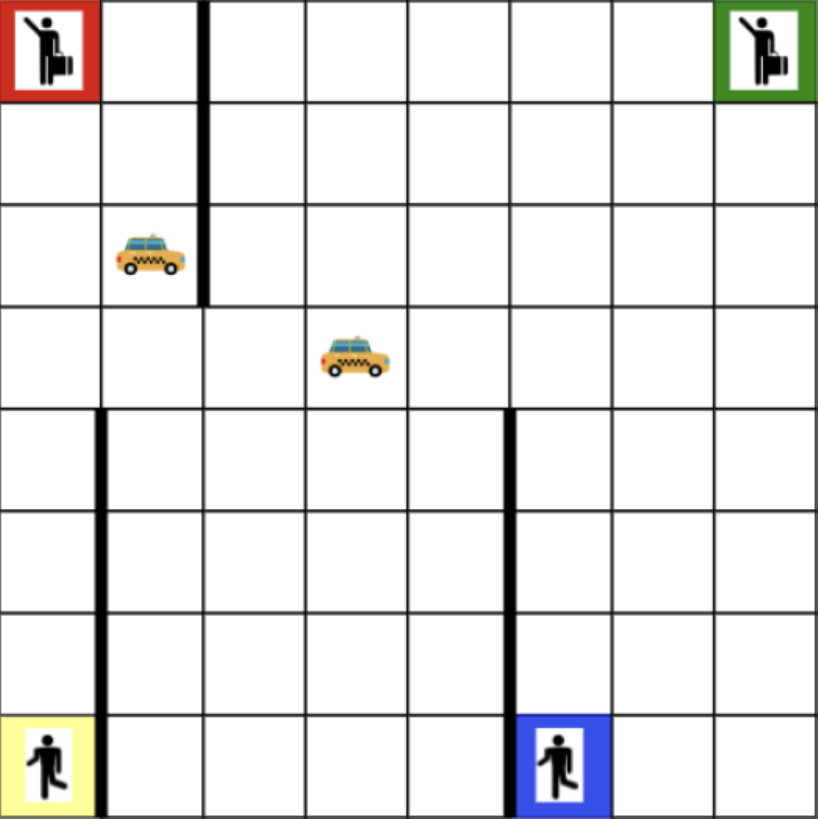}
        \caption{Multiagent Taxi}
      \end{subfigure}%
    \hfill
    \begin{subfigure}{0.28\linewidth}
        \includegraphics[width=\linewidth]{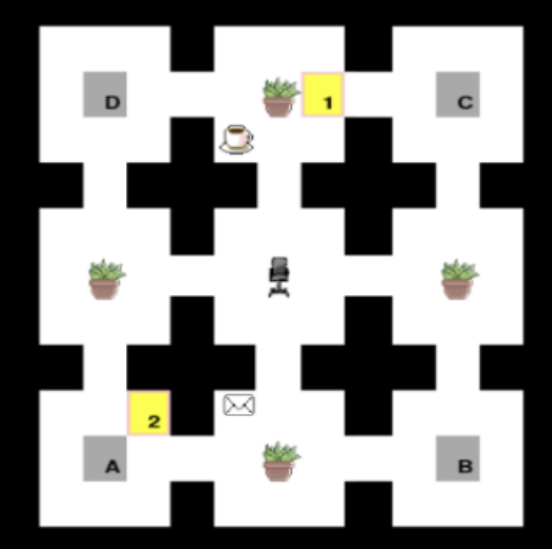}
        \caption{Multiagent Office}
    \end{subfigure}%
    \hfill
    \begin{subfigure}{0.28\linewidth}
    \includegraphics[width=\linewidth]{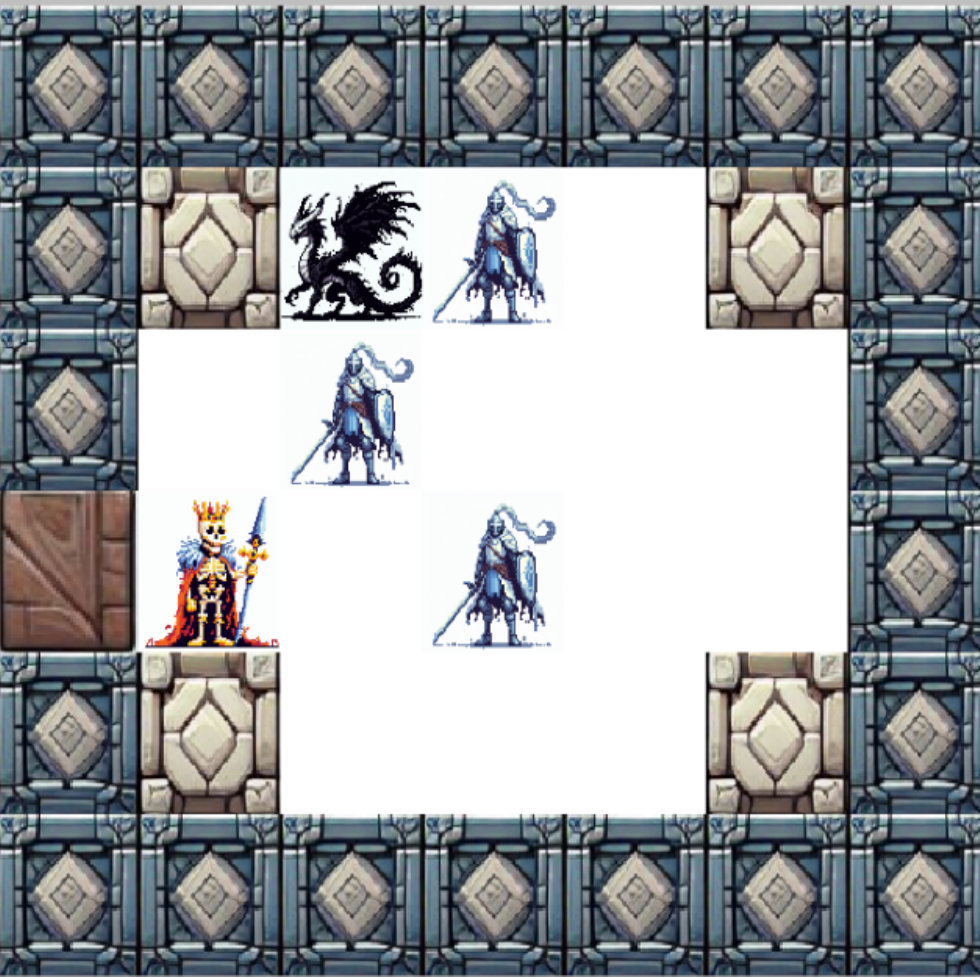}
        \caption{Multiagent Dungeon}
    \end{subfigure}%
    \hfill
    \caption{Relational Multiagent Domains}
    \label{fig:domains}
    \Description{Figures for the different relational multiagent domains (environments)}
\end{figure*}
\subsection{Multiagent Taxi}
The agent's goal in the Multiagent Taxi environment is to transport all the passengers to their respective destination locations, which involves picking up the passengers from one location and dropping them off at another. The observation, action, and rewards for the environment are defined as follows
\begin{enumerate}
    \item \textbf{Observation:} For each agent, the observation would contain the agent coordinates and the coordinates of the other agents. For each passenger, we would have its pickup location, drop location, as well as the taxi the passenger may be in. The pickup and drop locations are represented as one hot to indicate one of R, G, B, and Y
    \item \textbf{Actions:} The possible actions for an agent in a given state are
    \begin{enumerate}
        \item[] Move Up
        \item[] Move Down
        \item[] Move Left
        \item[] Move Right
        \item[] Pickup
        \item[] Drop
    \end{enumerate}
    \item \textbf{Rewards:} There is a -0.1 step cost associated with each step and a -1 cost associated with a no-move action. A reward of +20 will be provided for each pickup and drop. There is a high negative penalty of -100 for crashing into other agents followed by episode termination
    \item \textbf{Number of Agents:} While the environment can be extended to four agents, we considered only two agents for the experiments.
\end{enumerate}
\subsection{Multiagent Office}
The goal in the Multiagent Office environment is to finish all the designated tasks in the environment, which can be either visiting locations -- A, B, C, or D --, picking up coffee (C) or mail (M) and potentially dropping off to the office (O). 
The observations, actions, and rewards for the environment are defined as follows
\begin{enumerate}
    \item \textbf{Observation:} The observations for each agent would contain each agent's position, inventory for coffee and mail, and the facts of the environment state which is an indicator for each task possible in the environment and the value may indicate the agent which completed the task first
    \item \textbf{Actions:} The possible actions for an agent in a given state are
    \begin{enumerate}
        \item[] Move Up
        \item[] Move Down
        \item[] Move Left
        \item[] Move Right
    \end{enumerate}
    There are no separate pickup and drop actions, as an item like mail or coffee is automatically added to the inventory when the agent moves to its location in the environment. 
    \item \textbf{Rewards:} There is a -0.1 step cost associated with each step and a -1 cost for each invalid action. There is a -30 reward for bumping into other agents. There is a terminal reward of 100 provided to all the agents for completing the tasks in the episode
    \item \textbf{Number of Agents:} While the environment can be extended to accommodate $n$ agents, we only consider the case with two agents for the experiments.
\end{enumerate}
\subsection{Multiagent Dungeon}
The goal in the multiagent dungeon environment is to escape by unlocking the door using the keys collected by defeating all the enemies present in the environment. The enemies can be either skeletons with normal attack and hit points or dragons with high attack and normal hit points. The observation, actions, and rewards  for the environment are defined as follows
\begin{enumerate}
    \item \textbf{Observation:} The observation space for each agent would contain each agent's location and orientation, attack, hit points, and defense stance value. It also contains the location, hit points, key status (a binary indicating whether the key is in the door), and another binary value indicating if the agent is holding it. Finally, there is the door location as a value indicating the door-locked status
    \item \textbf{Actions:} The possible actions for an agent in the environment are
    \begin{enumerate}
        \item[] Move Up
        \item[] Move Down
        \item[] Move Left
        \item[] Move Right
        \item[] Attack
        \item[] Defend
        \item[] Pickup
        \item[] Unlock
    \end{enumerate}
    When an agent dies, it can no longer act in the environment. Moreover, any keys the agents collect are spawned at the agent's last location.
    \item \textbf{Rewards:} There is a -0.1 step cost associated with the environment. Moreover, there is a terminal reward of +100 provided to each agent, when the door is unlocked.  There are no rewards provided to any dead agents
    \item \textbf{Number of Agents:} While the domain can be expanded to handle $n$ agents, we consider only three agents for our experiments.
\end{enumerate}
\section{MaRePReL Abstractions}
MaRePReL employs an abstract state representation for its underlying RL policies, with the specific abstraction varying for each operator within a given domain. The D-FOCI statements corresponding to each operator are utilized to construct its abstract state and observation. Below, we present the abstract state used for each domain
\subsection{Multiagent Taxi}
The domain has two operators, \textbf{pickup} and \textbf{drop}. The abstract state representation considers both the passenger as well as the agent. The abstract representation for the operators is provided below
\begin{enumerate}
    \item \textbf{pickup}: The abstract state representation contains the locations (x, y) for all the agents, as well as the pickup location for the agent (one of R, G, B, and Y represented as a one-hot encoding in the observation), as well as the taxi\_id indicating the taxi the passenger is in (0 if the passenger is not in any taxi).
    \item \textbf{drop}: The abstract state representation contains the locations (x, y) for all the agents, as well as the taxi\_id for the taxi the passenger is in and the drop location for the agent (one of R, G, B, and Y represented as a one-hot encoding in the observation)
\end{enumerate}
\subsection{Multiagent Office}
The domain has two operators as well, \textbf{visitOrPickup} (picking up something is the same as moving to the place of the object in question), and \textbf{deliver}. The abstract state representation considers both the agents and the location (objects) in question
\begin{enumerate}
    \item \textbf{visitOrPickup}: The abstract state representation contains the location (x,y) of all the agents, the current agent's inventory (mail, coffee), as well as the corresponding fact (value indicating which agent performed the task).
    \item \textbf{deliver}: The abstract state representation contains the locations (x, y) of all the agents, the object inventory, and the fact corresponding to visit office
\end{enumerate}
\subsection{Multiagent Dungeon}
Like the other domains, Multiagent Dungeon has two operators as well \textbf{attackEnemy} and \textbf{getKeyinDoor}. The operators are parameterized by the agent as well as the enemy. The abstract representation for the operators is provided below
\begin{enumerate}
\item \textbf{attackEnemy}: The abstract state consists of all the agent locations (x, y), as well as the orientation, hp, and defense of the current agent. Furthermore, we include the current enemy's location (x, y) and hit points.
\item \textbf{getKeyInDoor}: The abstract state consists of all the agent's locations (x, y), the current enemy location (x,y), and its' corresponding key's status. 
\end{enumerate}
\section{Experiment Hyperparameters}
The learning algorithm for DQN (independent learners, parameter sharing, and sub-plan embeddings) and MaRePReL is DQN implemented using the ray rllib \footnote{https://docs.ray.io/en/latest/rllib/index.html} package in Python. The last baseline used is Q-Mix, which is also implemented using ray rllib. All the hyperparameters used except the ones described are the default parameters provided in ray for DQN or Q-Mix.
\subsection{Hyperparameters Dictionary}
The different unique parameters and the values assigned to them are described below
\begin{itemize}
    \item \textbf{model[epsilon\_timesteps]}: The  total number of time steps over which the epsilon is reduced from its initial value to the final value
    \item \textbf{model[final\_epsilon]}: The final epsilon value used after the initial reduction for the remainder of the run
    \item \textbf{model[target\_network\_update\_frequency]}: The number of steps after which we do a backward pass to update the target network parameters
    \item \textbf{model[fcnet\_hidden]}: It is a list that determines the number and size of the hidden layer for the model. A value of [256, 256] means that there are two hidden layers of size 256 each. 
    \item \textbf{model[fcnet\_activation]}: The activation function used in the DQN network
    \item \textbf{model[lstm\_cell\_size]}: The size of the lstm cell used for the model
    \item \textbf{model[max\_seq\_len]}: The maximum length used while using lstm in the algorithm
    \item \textbf{replay\_buffer[type]}: The type of replay buffer used by the DQN to store the trajectories. The one used throughout the experiments is a MultiAgentPrioritizedReplayBuffer. As the name indicates, it is a combination of Prioritized Replay Buffer and Multiagent Replay Buffer. The buffer stores the state-action transitions for the different agents, with the different transitions relative sample frequency determined by metrics such as the TD-error
    \item \textbf{replay\_buffer[capacity]}: The number of transitions that could be stored in the buffer
    \item \textbf{optimizer[alpha]}: The learning rate parameter for the network
    \item \textbf{optimizer[epsilon]}: For the optimizer, it is the value added to the gradients to prevent dividing by 0 errors.
    \item \textbf{double\_q}: A boolean flag indicating whether to use double DQN instead of the standard DQN algorithm.
    \item \textbf{n\_step}: The number of steps used for bootstrapping in multi-step TD learning
    \item \textbf{train\_batch\_size}: The size of the train batch used when training the model
    \item \textbf{lr}: The learning rate for the network
    \item \textbf{mixing\_embed\_dim}: The size of the hidden layer in the mixing dimension in the Q-mix network. Only applicable in the case of Q-Mix
\end{itemize}
\balance
\subsection{Hyperparameters for the different models and experiments}
\begin{table}[!ht]
\centering
\begin{tabular}{|l|l|}
\hline
 Hyperparameter        & Value \\\hline
    exploration\_config[epsilon\_timesteps]   & 10000\\
    exploration\_config[final\_epsilon] & 0.01 \\
    target\_network\_update\_frequency  & 2000  \\
    train batch size     & 128      \\
    replay\_buffer[type]    & MultiAgentPrioritized\\
    & ReplayBuffer      \\
    replay\_buffer[capacity]   & 300000      \\
    model[fcnet\_hidden] & [256, 256] \\
    model[fcnet\_activation] & relu \\
    double\_q    &  true      \\
    n\_step   &  4      \\\hline
\end{tabular}
\caption{\textbf{DQN (Independent Learners, Parameter Sharing, Plan Embeddings)} and \textbf{MaRePReL} lower level policies hyperparameters for MultiAgent Taxi}
\end{table}
\begin{table}[!ht]
    \centering
    \begin{tabular}{|l|l|}
    \hline
        Hyperparameter        & Value \\\hline
        exploration\_config[epsilon\_timesteps]   & 10000\\
        exploration\_config[final\_epsilon] & 0.01 \\
        model[fcnet\_hidden] & [256, 256] \\
        model[fcnet\_activation] & relu \\
        target\_network\_update\_frequency  & 500  \\
        train batch size     & 128      \\
        replay\_buffer\_[type]    &  MultiAgentPrioritized \\ 
        & ReplayBuffer      \\
        replay\_buffer\_config[capacity]   & 300000      \\
        double\_q    &  true      \\
        n\_step   &  4      \\\hline
    \end{tabular}
    \caption{\textbf{DQN (Independent Learners, Parameter Sharing, Plan Embeddings)} lower level policies hyperparameters for MultiAgent Office}
\end{table}
\begin{table}[!ht]
    \centering
    \begin{tabular}{|l|l|}
    \hline
        Hyperparameter        & Value \\\hline
        exploration\_config[epsilon\_timesteps]   & 10000\\
        exploration\_config[final\_epsilon] & 0.05 \\
        model[fcnet\_hidden] & [256, 256] \\
        model[fcnet\_activation] & relu \\
        target\_network\_update\_frequency  & 2000  \\
        train batch size     & 128      \\
        replay\_buffer\_[type]    & MultiAgentPrioritized \\
        & ReplayBuffer      \\
        lr & 0.001 \\
        replay\_buffer\_config[capacity]   & 300000      \\
        double\_q    &  true      \\
        n\_step   &  4      \\\hline
    \end{tabular}
    \caption{\textbf{DQN (Independent Learners, Parameter Sharing, Plan Embeddings)} lower level policies hyperparameters for MultiAgent Dungeon}
\end{table}
\newpage\noindent
\begin{table}[!ht]
\centering
\begin{tabular}{|l|l|}
\hline
    Hyperparameter        & Value \\\hline
    mixing\_embed\_dim & 256 \\
    optimizer[alpha] & 0.99 \\
    optimizer[epsilon] & 1e-5 \\
    model[lstm\_cell\_size] & 64 \\
    model[max\_seq\_len] & 200 \\
    model[fcnet\_hidden] & [256, 256] \\
    model[fcnet\_activation] & relu \\
    exploration\_config[epsilon\_timesteps]   & 10000\\
    exploration\_config[final\_epsilon] & 0.01 \\
    target\_network\_update\_frequency  & 2000  \\
    train batch size     & 128      \\
    replay\_buffer\_[type]    & ReplayBuffer      \\
    replay\_buffer\_config[capacity]   & 100000      \\
    double\_q    &  true      \\\hline
\end{tabular}
  \caption{\textbf{QMIX} hyperparameters for Multiagent Taxi and Multiagent Office}
\end{table}

\end{document}